\documentclass[12pt]{article}

\usepackage{scicite}

\usepackage{times}

\usepackage{graphicx}

\newenvironment{sciabstract}{%
\begin{quote} \bf}
{\end{quote}}

\title{Microlensing events indicate that super-Earth exoplanets are common in Jupiter-like orbits} % 91 characters with spaces

\author
{Weicheng Zang$^{1,2\ast}$, Youn Kil Jung$^{3,4\ast}$, Jennifer C. Yee$^{2}$, Kyu-Ha Hwang$^{3}$, \and
Hongjing Yang$^{1,5}$, Andrzej Udalski$^{6}$, Takahiro Sumi$^{7}$, Andrew Gould$^{8,9}$, Shude Mao$^{5,1\ast}$, \and
Michael D. Albrow$^{10}$, Sun-Ju Chung$^{3}$, Cheongho Han$^{11}$, Yoon-Hyun Ryu$^{3}$, \and
In-Gu Shin$^{2}$, Yossi Shvartzvald$^{12}$, Sang-Mok Cha$^{3,13}$, Dong-Jin Kim$^{3}$,\and 
Hyoun-Woo Kim$^{3,11}$, Seung-Lee Kim$^{3}$, Chung-Uk Lee$^{3\ast}$, Dong-Joo Lee$^{3}$, \and
Yongseok Lee$^{3,13}$, Byeong-Gon Park$^{14}$, Richard W. Pogge$^{9,15}$,
Xiangyu Zhang$^{8}$, \and
Renkun Kuang$^{1,16}$, Hanyue Wang$^{17}$, Jiyuan Zhang$^{1}$, 
Zhecheng Hu$^{1}$, Wei Zhu$^{1}$, \and
Przemek Mr\'{o}z$^{6}$, Jan Skowron$^{6}$, Rados{\l}aw Poleski$^{6}$, Micha{\l} K. Szyma\'{n}ski$^{6}$, \and
Igor Soszy\'{n}ski$^{6}$, Pawe{\l} Pietrukowicz$^{6}$, Szymon Koz{\l}owski$^{6}$, Krzysztof Ulaczyk$^{18}$, \and
Krzysztof A. Rybicki$^{6,12}$, Patryk Iwanek$^{6}$, Marcin Wrona$^{6,19}$, Mariusz Gromadzki$^{6}$, \and
Fumio Abe$^{20}$, Richard Barry$^{21}$, David P. Bennett$^{21,22}$, Aparna Bhattacharya$^{21,22}$, \and
Ian A. Bond$^{23}$, Hirosane Fujii$^{20,24}$, Akihiko Fukui$^{25,26}$, Ryusei Hamada$^{7}$, Yuki Hirao$^{27}$, \and
Stela Ishitani Silva$^{21}$, Yoshitaka Itow$^{20}$, Rintaro Kirikawa$^{7}$, Naoki Koshimoto$^{7}$, \and
Yutaka Matsubara$^{20}$, Shota Miyazaki$^{28}$, Yasushi Muraki$^{20}$, Greg Olmschenk$^{21}$, \and
Cl\'{e}ment Ranc$^{29}$, Nicholas J. Rattenbury$^{30}$, Yuki Satoh$^{7,31}$, Daisuke Suzuki$^{7}$, \and
Mio Tomoyoshi$^{7}$,  Paul J. Tristram$^{32}$, Aikaterini Vandorou$^{22,23}$, \and
Hibiki Yama$^{7}$, and Kansuke Yamashita$^{7}$\and \vspace{-3cm} 
\\ 
\small$^{1}$Department of Astronomy, Tsinghua University; Beijing, China\\
\small$^{2}$Solar, Stellar, and Planetary Physics Division, Center for Astrophysics $|$ Harvard Smithsonian; Cambridge, MA, USA\\
\small $^{3}$Optical Astronomy Division, Korea Astronomy and Space Science Institute; Daejeon, Republic of Korea \\
\small $^{4}$Department of Astronomy and Space Science, National University of Science and Technology, Korea; \\ \small  Daejeon, Republic of Korea\\
\small $^{5}$Department of Astronomy, Westlake University; Hangzhou 310030, China\\
\small $^{6}$Astronomical Observatory, University of Warsaw; Warszawa, Poland\\
\small $^{7}$Department of Earth and Space Science, Graduate School of Science, Osaka University; Osaka, Japan\\
\small $^{8}$Galaxies and Cosmology, Max-Planck-Institute for Astronomy; Heidelberg, Germany \\
\small $^{9}$Department of Astronomy, Ohio State University; Columbus, OH, USA \\
\small $^{10}$Department of Physics and Astronomy, University of Canterbury; Christchurch, New Zealand\\
\small $^{11}$Department of Physics, Chungbuk National University; Cheongju, Republic of Korea\\
\small $^{12}$Department of Particle Physics and Astrophysics, Weizmann Institute of Science; Rehovot, Israel \\
\small $^{13}$School of Space Research, Kyung Hee University; Kyeonggi, Republic of Korea \\
\small $^{14}$Center for Large Telescopes, Korea Astronomy and Space Science Institute; Daejeon, Republic of Korea\\
\small $^{15}$Center for Cosmology and AstroParticle Physics, Ohio State University; Columbus, OH, USA\\
\small $^{16}$Department of Engineering Physics, Tsinghua University; Beijing, China\\
\small $^{17}$Astronomy Department, Harvard University; Cambridge, MA, USA\\
\small $^{18}$Department of Physics, University of Warwick; Coventry, UK\\
\small $^{19}$ Department of Astrophysics and Planetary Sciences, Villanova University; 
\small Villanova, PA 19085, USA\\
\small $^{20}$Institute for Space-Earth Environmental Research, Nagoya University; Nagoya, Japan\\
\small $^{21}$Exoplanets and Stellar Astrophysics Laboratory, NASA Goddard Space Flight Center; Greenbelt, MD, 20771, USA\\
\small $^{22}$Department of Astronomy, University of Maryland; College Park, MD, USA\\
\small $^{23}$Institute of Natural and Mathematical Sciences, Massey University; Auckland, New Zealand \\
\small $^{24}$ Department of Earth and Space Science, Graduate School of Science, Osaka University; Osaka, Japan\\
\small $^{25}$Department of Earth and Planetary Science, Graduate School of Science, The University of Tokyo; Tokyo, Japan\\
\small $^{26}$Instituto de Astrof\'isica de Canarias; La Laguna, Tenerife, Spain \\ % No department
\small $^{27}$Institute of Astronomy, Graduate School of Science, The University of Tokyo; Tokyo, Japan\\
\small $^{28}$Institute of Space and Astronautical Science, Japan Aerospace Exploration Agency; Kanagawa, Japan\\
\small $^{29}$
Institut d'Astrophysique de Paris, Sorbonne Universit\'{e},\\
    \small Centre National de la Recherche Scientifique (CNRS); 75014 Paris, France.\\
\small $^{30}$Department of Physics, University of Auckland; Auckland, New Zealand \\
\small $^{31}$College of Science and Engineering, Kanto Gakuin University; Yokohama, Japan\\
\small $^{32}$Mount John Observatory, University of Canterbury; Lake Tekapo, New Zealand\\
\\
$^\ast$Corresponding author e-mail:  weicheng.zang@cfa.harvard.edu, \\
ykjung21@kasi.re.kr, leecu@kasi.re.kr,  shude.mao@westlake.edu.cn.
}

\date{}

\begin{document} 


\baselineskip24pt


\maketitle 


\begin{sciabstract}
Exoplanets classified as super-Earths 
are commonly observed on short period orbits, close to their host stars, but their abundance on wider orbits is poorly constrained. Gravitational microlensing is sensitive to exoplanets on wide orbits.
We observed the microlensing event OGLE-2016-BLG-0007, which indicates an exoplanet
with a planet-to-star mass ratio roughly double the Earth-Sun mass-ratio, on an orbit longer than Saturn's.
We combine this event with a larger sample from a microlensing survey to determine the distribution of mass ratios for planets. 
%on wide orbits.
We infer there 
are $\sim 0.35$ super-Earth planets per star on Jupiter-like orbits. 
The observations are most consistent with a bimodal distribution, with separate peaks for super-Earths and Jupiters. We suggest this reflects differences their formation processes.
\end{sciabstract}

% In setting up this template for *Science* papers, we've used both
% the \section* command and the \paragraph* command for topical
% divisions.  Which you use will of course depend on the type of paper
% you're writing.  Review Articles tend to have displayed headings, for
% which \section* is more appropriate; Research Articles, when they have
% formal topical divisions at all, tend to signal them with bold text
% that runs into the paragraph, for which \paragraph* is the right
% choice.  Either way, use the asterisk (*) modifier, as shown, to

Information about the formation and evolution of planetary systems is encoded in the distribution
of exoplanet masses and orbital separations. Many exoplanets have been 
observed on
short-period ($P < 1~\mathrm{yr}$) orbits, including large numbers
of super-Earths, 
\cite{Borucki11, Mayor11, Wittenmyer11, Bonfils13}
which are planets larger than Earth, but smaller than Neptune. These terms are qualitative, so they can be defined as a range of either planetary radii, planetary masses, or planet-star mass ratios depending on the observing technique.
For longer-period orbits  ($P > 1~\mathrm{yr}$), 
only the frequency of 
larger planets, gas and ice giants, 
has been determined
\cite{Suzuki16, Shvartzvald16, Poleski21}. 
The population of smaller planets, including those we classify as Earth and super-Earth planets because the logarithms of their mass ratios ($q$) are $\log q < -4.5$, on orbits with $P > 1~\mathrm{yr}$ is poorly constrained.
Microlensing is sensitive to
such planets 
(Figure 1).

Gravitational microlensing occurs when a foreground object (referred to as the lens) passes between an observer and a background object (the source). This focuses the light from the source, causing a transient apparent increase in the source brightness as a function of time (its light curve).
For microlensing events in which both the lens and source are stars, this brightening typically lasts several months. If the lens star is orbited by a planet with a suitable alignment, an additional brightening is superimposed on the light curve, lasting days or hours.
Super-Earth planets are challenging to detect with microlensing
because 
they often cause only a small brightening of the light curve
for only a few hours. 
Light curves measured by a single observatory 
will have gaps 
due to the day-night cycle 
or poor weather.
Therefore, detecting super-Earth planets using microlensing usually requires combining data from multiple observatories to provide a continuous light curve.

We analyze data from the Korea Microlensing Telescope Network
(KMTNet), which has telescopes at three different sites \cite{Kim16_KMTNet}.
Each telescope provides 2 to 40 observations per night of the same target region, so the network provides densely sampled light curves \cite{KimKim18_EF,Kim18EF,Kim18_AF}. 
Microlensing events with planets are identified in the KMTNet data by a semi-automated pipeline called \textsc{AnomalyFinder} \cite{Zang21AF1, Zang22AF4}.

\section*{A long-period super-Earth planet}

The microlensing event OGLE-2016-BLG-0007 was initially identified as an ongoing microlensing event by the Optical Gravitational Lensing Experiment (OGLE) on 2016 February 10 \cite{Udalski1994,Udalski2003}. It was also observed by KMTNet and the Microlensing Observations in Astrophysics (MOA) collaboration. 
Our analysis of the KMTNet data with \textsc{AnomalyFinder} \footnote{Materials and methods are available as supplementary materials.} identified an additional bump in the light curve. Figure 2 shows the light curve including data from all three groups.
The bump has a low-magnification, $A_{\rm bump} \simeq 1.09$, peaking at time $t_{\rm bump} \sim$ 2016 September 25 UT 00 $ \sim 7656.5$  HJD$^\prime$ (where HJD$^\prime$ is the heliocentric Julian date minus 2,450,000), which lasts for $\Delta t_{\rm bump} \sim 2.1~\mathrm{days}$. 
The bump is superimposed on a standard stellar microlensing curve with parameters
$(t_0, u_0, t_{\rm E}) = (7498.4$ $\sim$ 2016 April 19 UT 21:30,  $1.253, 73.8~{\rm days})$, where
$(t_0, u_0, t_{\rm E})$ are the time of the peak of the event, the impact parameter between the source and lens stars, and the timescale of the event.
We investigate whether the bump is caused by a small planet orbiting the lens star.

There are three possible configurations that could produce this light curve (Fig. S2). We investigate each, finding that the most plausible solution (the Wide model) is a planet located more than one Einstein radius ($\theta_{\rm E}$, the characteristic spatial scale of the event) from the lens star.
The two other possible configurations involve either
a larger planet (higher $q$) located less than one Einstein radius from its star or 
a second (much fainter) source star rather than a planet \cite{Gaudi98}. 
We modeled both of these alternate possibilities but rule them out at high confidence ($\Delta\chi^2 = 38.5$ and 145.6, respectively, see Table S1).

 The Wide model indicates the separation $s$ between the host star and the planet is $s =2.83\pm0.01~\theta_{\rm E}$ (values and uncertainties are medians and 68\% confidence ranges respectively, unless otherwise stated) and the mass ratio between the planet and its star $q$ is $q = 6.79 \times 10^{-6}$ ($\log q = -5.17\pm0.13$).
 This mass ratio is roughly twice that of the mass ratio between the Earth and Sun ($q_{\rm Earth} \equiv M_{\rm Earth} / M_{\rm Sun} = 3 \times 10^{-6}$, where $M_{\rm Earth}$ and $M_{\rm Sun}$ are the masses of Earth and the Sun, respectively). This makes OGLE-2016-BLG-0007Lb (where ``L" indicates the lens star and ``b" indicates the planet orbiting that star) a super-Earth planet.

We combine the microlensing parameters with a model of the distribution of stars in the Milky Way galaxy to estimate the likely lens
mass $0.59_{-0.30}^{+0.41}~ M_{\rm Sun}$ and distance from the Sun $4.3_{-1.4}^{+1.1}~\mathrm{kiloparsecs~(kpc)}$.
These values imply a planet with a mass of $m_{\rm p} = 1.32_{-0.67}^{+0.91}~ M_{\rm Earth}$ on an orbit with semi-major axis $a = 10.1_{-3.4}^{+3.8}~\mathrm{au}$ (where $1~\mathrm{au}$ is approximately the distance from the Earth to the Sun) and period $P= 39_{-9}^{+21}~\mathrm{yr}$ (Table 1).

These properties correspond to a super-Earth planet with an orbital period longer than Saturn's. 
Such planets are 
commonly produced by
population synthesis models \cite{Burn21_BERN4}, but they are absent in previous observations of normal (main sequence) stars (Figure 1). The only other planet similar to OGLE-2016-BLG-0007Lb was found in orbit around a pulsar \cite{StarovoitRodin17}. Planets known to orbit pulsars are excluded from the Figure because the supernova that created the pulsar is assumed to have strongly affected its local planetary system, and the detected pulsar planets may even have formed after the explosion \cite{PatrunoKama17}. 

\section*{Mass-Ratio Distribution from Microlensing}

To investigate the population of planets detected using microlensing, we combine OGLE-2016-BLG-0007Lb with the other planets identified in the KMTNet survey using \textsc{AnomalyFinder}.
Our dataset consists of observations from KMTNet seasons 2016 to 2019.
All planets with $\log q > -4$ are from the 2018 and 2019 KMTNet seasons \cite{Zang22AF4,Gould22AF5,Jung22AF6,Jung23AF8}. 
Planets with $\log q < -4$ are 
rarer, so we include planets identified in  
the 2016 to 2019 KMTNet seasons
\cite{Zang21AF1,Hwang22AF2,Zang23AF7}, such as OGLE-2016-BLG-0007Lb. 
In total, our
sample consists of 63
planets identified in 60 microlensing events observed with KMTNet. 

To determine the sensitivity of
the \textsc{AnomalyFinder} algorithm and KMTNet observations, we inject
simulated planet signals into measured
light curves from the 2018 and 2019 seasons, then determine the fraction of those that are recovered by the algorithm.
The sensitivity functions were calculated separately for the 2018 and 2019 seasons; we find they
are very similar, especially at $\log q < -4$ (Figure S1).
KMTNet maintained the same observational setup from 2016 to 2019 with only minor changes due to routine maintenance, so we assume 2016 and 2017 observations have a sensitivity function equal to the average of the 2018 and 2019 sensitivity functions.

We 
correct the number of planet detections using the
sensitivity functions for each year to determine the underlying planet distribution. 
This determines the planet frequency distribution from $\log q = -5.2$ to $\log q = -1.5$. The lower limit is set by OGLE-2016-BLG-0007Lb, which has a mass-ratio of $\log q = 5.17$, the lowest in the sample.
Over this range ($-5.2 < \log q < -1.5$), the total planet frequency is $0.65^{+0.35}_{-0.15}$ planets per star per dex (where `dex' is one decade in $\log s$). 
For super-Earths
, which we assume occupy the range
$-5.2 < \log q < -4.5$ (the upper limit would correspond to a $\sim 10~ M_{\rm Earth}$ planet orbiting a $1~ M_{\rm Sun}$ host), there are $0.35^{+0.34}_{-0.14}$ planets per star per dex.

Fig. 3A shows the distribution of logarithmic mass ratios $\log q$ of the planet detections and the planet distribution corrected by the sensitivity function.  The bin size changes at $\log q = -4$ because four seasons (2016-2019) of detections are considered for $\log q < -4$, but above that, we use only the 2018 and 2019 seasons. For $-6.0 < \log{q} < -5.25$, we show the upper limits on the planet frequencies from Poisson statistics. If we were trying to demonstrate that the non-detections are significant, $3\sigma$ limits would be appropriate. We show $1\sigma$ limits to demonstrate the opposite: a substantial population of planets could be consistent with the non-detections.

We fit the distribution using a single power-law function,
\begin{equation}
\frac{dN}{d\log q} = 0.18 \pm 0.03 \left(\frac{q} {10^{-4}}\right)^{-0.55\pm0.05}
\end{equation}
where $N$ is the number of planets per star,
finding it qualitatively reproduces the data (see Fig 3).

Fig. 3B shows the cumulative distribution of individual detections as a function of $\log q$. In this Figure, the power-law model, rather than the data, is corrected by the sensitivity function. The slight break in the model distribution at $\log q = -4$ reflects the change in the number of observational seasons used in the analysis above and below this point.

Planet formation theory does not predict a single power law distribution from super-Earths to gas giants.
For example, in the core-accretion planet formation scenario, runaway gas accretion occurs once a proto-planet exceeds a critical mass, causing them
 to rapidly grow from tens to hundreds of Earth masses \cite{Pollack96,PapaloizouNelson05,Coleman17}. 
This process would produce two separate populations of planets in different mass regimes, one at $\sim10~ M_{\rm Earth}$ and one above $\sim100~ M_{\rm Earth}$ with a deficit between them.
The alternative gravitational instability scenario produces gas
giant planets
 via direct collapse of a proto-planetary disk \cite{Boss97} 
 but does not produce lower-mass rocky planets, which must form via a different mechanism. This would also produce a population of gas giants (with masses above $\sim 100~ M_{\rm Earth}$) and a separate population of less massive planets.

We therefore investigated whether there is evidence for two populations in our derived mass ratio distribution. There is a sharp change in the slope of the cumulative mass-ratio distribution (Fig. 3B) at $\log q = [-3.6, -3]$ (also visible as a slight dip at $\log q = [-3.5, -3]$ in Fig. 3A).  Because $\log q = -3.2$ for a $\sim 100~ M_{\rm Earth}$ planet orbiting a $0.5~ M_{\rm Sun}$ star, this is approximately where we expect the division between planets that have and have not undergone runaway gas accretion. In addition, there appear to be other changes in the slope of the cumulative mass ratio distribution
at both the high ($\log q > -2$) and low ($\log q < -5$) ends (Fig. 3B). 
The power-law model predicts four planets with $-6 < \log q < -5$ in the KMTNet sample, but only one has been detected, which has a probability of  9\% assuming a Poisson distribution.
We reproduce these features by fitting a double-Gaussian function of $\log q$ (also shown in Fig. 3):
\begin{equation}
\frac{dN}{d\log q} = \mathcal{A}_1 \times 10^{ n_1 (\log q - \log q_{\rm peak, 1})^2} 
+ \mathcal{A}_2 \times 10^{n_2 (\log q - \log q_{\rm peak, 2})^2}.
\end{equation}
which has peaks at $\log q_{\rm peak, 1} = -4.7^{+0.2}_{-0.3} = \log(7.4 \times q_{\rm Earth})$ and $\log q_{\rm peak, 2} = -2.6\pm0.1 = \log(770 \times q_{\rm Earth})$, 
amplitudes $\mathcal{A}_1 = 0.52\pm0.13$ and $\mathcal{A}_2 = 0.058 \pm 0.013$, and $n_1 = -0.73\pm0.35$ and $n_2 = -1.8\pm0.6$, where $n_1^{-1/2}$ and $n_2^{-1/2}$ are the effective widths of the Gaussians. 

Quantitatively, the double-Gaussian is a better fit than the single-power law by $-2\Delta\ln {\cal L} = 22.6$, where $-2\Delta\ln {\cal L}$ is -2 times the difference in the log likelihoods of the models.
We calculated the false alarm probability $p$ is $1.60\times 10^{-4}$ for observing a $-2\Delta\ln {\cal L}=22.6$ preference for a double-Gaussian model if the true distribution is the power law described above. This indicates statistical preference for the double-Gaussian model over the power law model, despite the increased number of free parameters. We also regard the double-Gaussian model as more consistent with theoretical predictions, because it has $\log q$ peaks in the expected ranges of two classes of planets
formed by runaway gas accretion. 

Integrating each Gaussian separately from $\log q = -5.2$ to $-1.5$ yields $0.57\pm0.13$ planets per star per dex in the lower mass ratio Gaussian and $0.053\pm0.013$ planets per star per dex in the higher mass-ratio Gaussian.

\section*{Comparison to Previous Work}

Independent of the specific functional form of the mass-ratio distribution, our results indicate an abundance of super-Earths in Jupiter-like orbits. 
This was not evident in previous studies of the microlensing mass-ratio distribution that
found a mass-ratio distribution described by a broken-power-law 
with a sharp break around $\log q \sim -4$ \cite{Suzuki16,Udalski18,Jung19_0165}.
In contrast, we find no evidence for a break at either $\log q =-3.77$ or $-4.26$ as found in those studies. 
The change in slope we identify in the cumulative mass-ratio distribution did not previously appear 
\cite{Suzuki16}. 
Figure S9 compares the two cumulative distributions; the previous work found an over-abundance of planets 
from $\log q = -3.5$ to $-3.0$, relative to a smooth distribution, which contributed to the previous preference for a broken power-law distribution \cite{Suzuki16}.
Differences in the methodology may contribute to differences in the inferred distributions (see Supplementary Text).

Our mass-ratio distribution is more consistent with a previous microlensing study of only
nine planets, the smallest of which had $\log q = -4.32$  \cite{Shvartzvald16}. That also did not find
a change in slope in the cumulative mass-ratio distribution 
in the range $\log q = [-3.6, -3]$, but nor was there an over-abundance of planets in that range. We attribute this to the low number of planets included in that study.
The power-law model that study fitted to the data in the range $-4.9 < \log q <-1.4$ 
is consistent with our power law model, though had larger uncertainties \cite{Shvartzvald16}.

Another study of six microlensing planets spanning log q = [-4.5, -2], found three of those planets were in the range $\log q = [-3.6, -3]$ \cite{Gould10}.
This is a greater proportion than we find, but we assess the small number of planets included in that study or differences in methodology could explain the tension between the results (see Supplementary Text).

Comparisons between microlensing results and planets detected using the alternative radial velocity technique are complicated because they are sensitive to planets on different orbits \cite{ClantonGaudi16}, measure different quantities (mass, $m_{\rm p}$, or $m_{\rm p} \sin i$, where $i$ is the inclination of the orbit, rather than mass ratio), and  the properties of most microlensing host stars are not measured whereas radial velocity studies tend to group their results by host star spectral type. We nevertheless compared our results to those from radial velocity studies (see Supplementary Text).

\section*{Implications for planet formation}

We compare our results to predictions from theoretical simulations of planet formation. A direct, quantitative comparison is
challenging because the 
host stars of microlensing planets span a wide range of masses, 
and the mass of any particular host star is usually unknown or poorly constrained.
Previous comparisons  \cite{Suzuki18} 
have shown that the population of microlensing planets is not consistent with the theoretical predictions, especially when the simulations include the effect of planet migration \cite{GoldreichTremaine80, LinPapaloizou86}. 

We qualitatively compare our mass-ratio distribution to predictions from simulations
[\cite{Suzuki18}, their figure 2]; we find that they are not 
consistent if the simulations include planet migration. However, alternative models \cite{IdaLin04, Ida05} without migration produce two distinct populations of planets, which is qualitatively similar to our derived distribution, although the locations and heights of the peaks differ from our fitted values. 
Other models \cite{Alibert05, Mordasini12} without migration are also qualitatively consistent with our observations, although updated versions of those models do not form any giant planets around stars $< 0.5 M_{\rm Sun}$ \cite{Burn21_BERN4}. 
Because any type of star can produce a microlensing event, many of the events are expected to be produced by the most abundant type of star, those $< 0.5 M_{\rm Sun}$.
We derive a giant planet ($-4 < \log q < -1.5$) frequency of $12.3 ^{+3.9}_{-2.6}\%$. This can only be consistent with the updated versions of the models if the great majority microlensing events with giant planets are produced by host stars with masses $M_{\rm host} = 0.7 M_{\rm Sun}$ and above.

We conclude that the two separate populations we identify in the mass ratio distribution indicate
differentiation during the planet formation processes.  
This could be explained by a single planet formation scenario in which the two populations are produced by runaway gas accretion for planets above a mass threshold. 
Alternatively, the two populations could be produced by different formation mechanisms:
accretion and
gravitational instability \cite{Boss23}. 

\clearpage
    \renewcommand\arraystretch{1.25}
    \begin{center}
    \begin{tabular}{l l r}
    \multicolumn{3}{c}{\bf Table 1. Derived system parameters for OGLE-2016-BLG-0007Lb}\\
    \hline
    Parameter & Symbol (units) & Value ($1\sigma~$ uncertainty) \\
    \hline
    Planet/host mass ratio  & $q~(10^{-6})$ & $6.9_{-1.9}^{+2.3}$ \\
    \hline
    Planet mass & $m_{\rm p}~(M_{\rm Earth})$ & $1.32_{-0.67}^{+0.91}$ \\
    \hline
    Host mass & $M_{\rm host}~(M_{\rm Sun})$ & $0.59_{-0.30}^{+0.41}$ \\
    \hline
    Lens distance & $D_{\rm L}$ (kpc) & $4.3_{-1.4}^{+1.1}$ \\
    \hline
    Two-dimensional host–planet separation & $a_{\bot}$ (au) & $8.9_{-2.9}^{+2.3}$ \\
    \hline
    Semi-major axis & $a$ (au) & $10.1_{-3.4}^{+3.8}$ \\
    \hline
    Planet orbital period & $P$ (yr) & $39_{-9}^{+21}$ \\
    \hline
    \end{tabular}
    \end{center}

\clearpage
\clearpage
\begin{figure}[htb] 
    \centering
    \includegraphics[width=\textwidth]{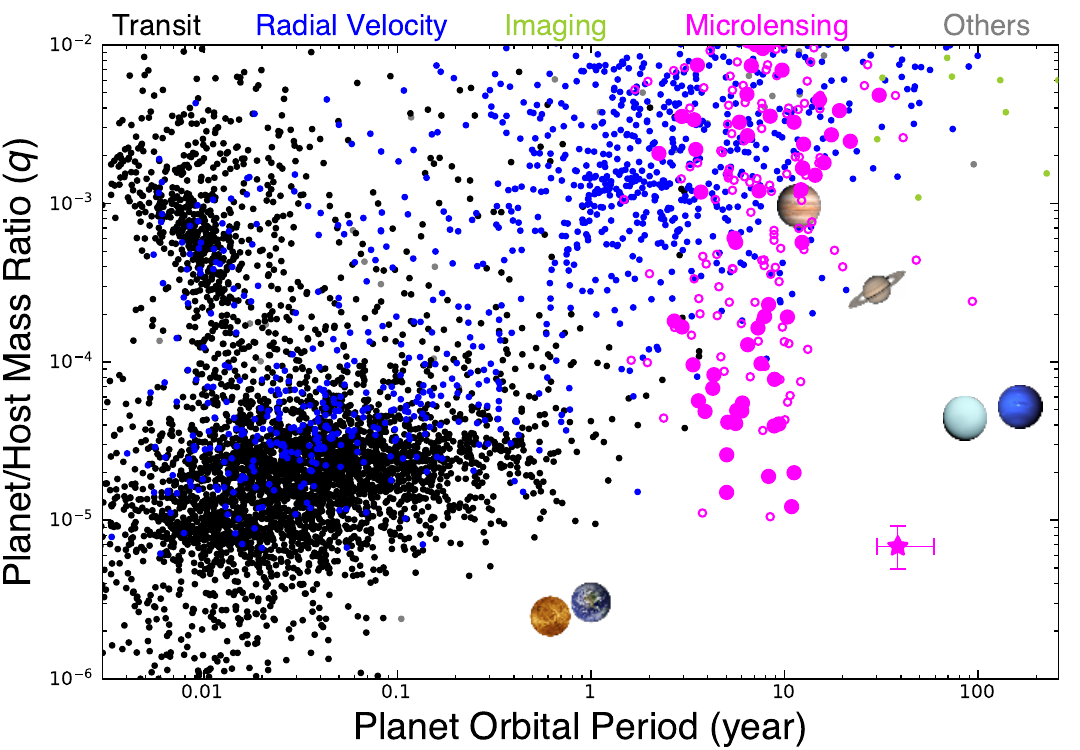}
    \caption{
    {\bf Periods and mass ratios for planets detected using different methods.}
    Data points show all exoplanets catalogued on 30 July 2023 \cite{NEA} and are color-coded by detection method (see legend). The magenta star indicates OGLE-2016-BLG-0007Lb (error bars show 68\% confidence intervals for this planet, but are omitted from all other planets).
 Colored images show six Solar System planets: Venus, Earth, Jupiter, Saturn, Uranus and Neptune.
Filled magenta points are our collated sample of planets from the KMTNet microlensing survey (see text); empty magenta circles are other planets detected with the microlensing technique.
 The microlensing detections include planets with mass-ratios corresponding to super-Earths
with Jupiter-like orbital periods.}
\end{figure}

\begin{figure}[htb] 
    \centering
    \includegraphics[height=0.8\textheight]{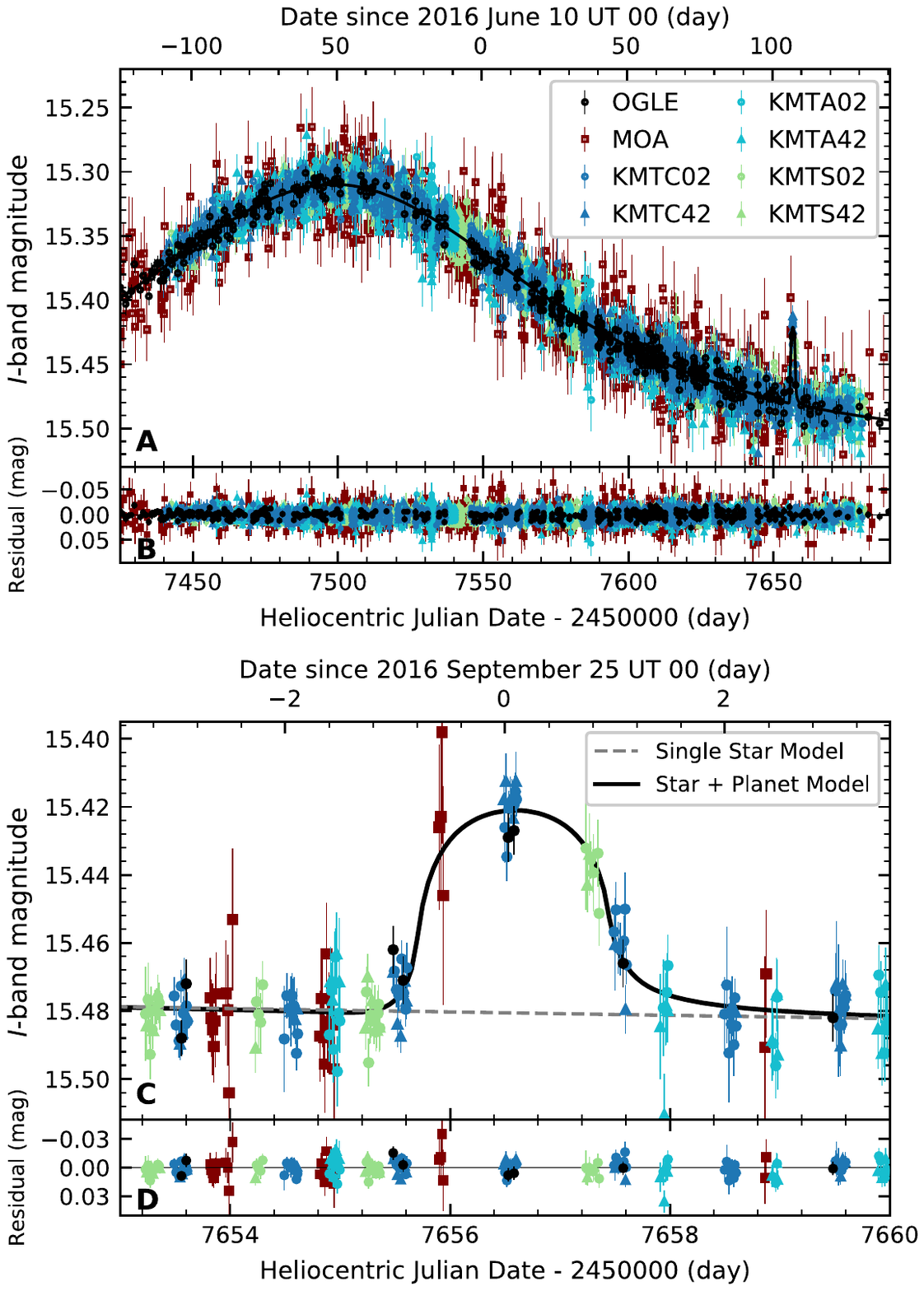}
    \caption{{\bf Light curve of OGLE-2016-BLG-0007.} Data points are color-coded by source (see legend) and shown with $1\sigma$ error bars. The black-solid and gray-dashed lines shows the best-fitting model light curves (see legend). {\bf A:} All observations of the event taken in 2016. {\bf C:} Zoom highlighting the low-amplitude, short-duration anomaly in OGLE-2016-BLG-0007, indicating a planet with an extremely small mass ratio of $\log q \sim -5$. {\bf B} and {\bf D:} residuals from the best-fitting planetary model. } 
\end{figure}

\begin{figure}
	\centering
	\includegraphics[height=0.3\textheight]{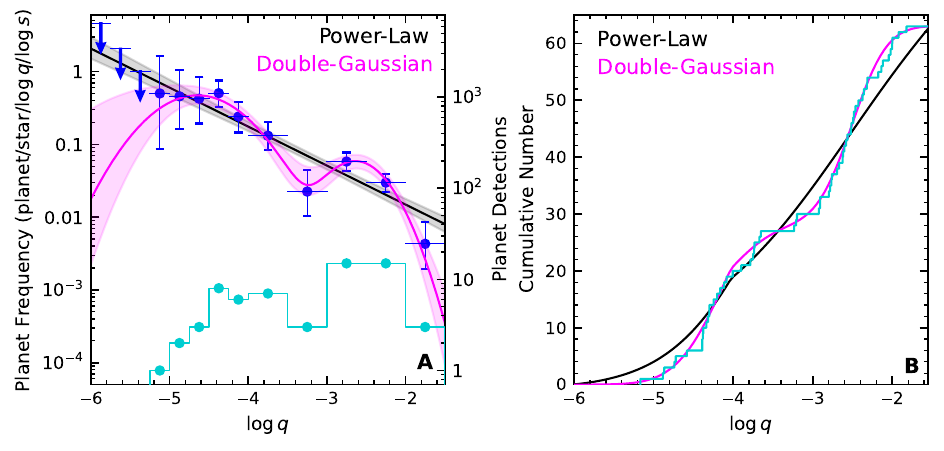}
	\caption{ {\bf Mass ratio distributions of microlensing planets from our KMTNet sample. A:} The observed number of planet detections (cyan; right axis) and the derived planet frequency distribution (blue; left axis) after correction by the sensitivity function (Fig S1.). The horizontal lines correspond to the width of the histogram bins.
	The error bars are $1\sigma$ Poisson uncertainties; the upper limits are also shown at $1\sigma$. Overlain are two models fitted to the data: a single power law (black line) and a double Gaussian (magenta line); shaded regions indicate 68\% confidence intervals.
	{\bf B:}
	The same two fitted models multiplied by the sensitivity function and expressed as a cumulative distribution. The cyan histogram is the cumulative distribution from the KMTNet sample. 
The power law model predicts more planets in the range $-3.5 < \log q < -3$ than were detected. The double Gaussian model is also a better match to the data at both high and low $q$ ends of the distribution.}
\end{figure}

\clearpage
% Your references go at the end of the main text, and before the
% figures.  For this document we've used BibTeX, the .bib file
% scibib.bib, and the .bst file Science.bst.  The package scicite.sty
% was included to format the reference numbers according to *Science*
% style.

%BibTeX users: After compilation, comment out the following two lines and paste in
% the generated .bbl file. 

%\bibliography{Zang.bib}
%\bibliographystyle{sciencemag}

%\bibliography{combined_final.bib}

%\bibliographystyle{Science}

\clearpage
\section*{Acknowledgments}
% Collaboration acknowledgements
The KMTNet system is
operated by the Korea Astronomy and Space Science Institute
      (KASI) at three host sites of CTIO in Chile, SAAO in South
      Africa, and SSO in Australia. Data transfer from the host site to
      KASI was supported by the Korea Research Environment
      Open NETwork (KREONET). 
% General resource acknowledgements
We acknowledge the NASA Exoplanet Archive, which is operated by the California Institute of Technology, under contract with the National Aeronautics and Space Administration under the Exoplanet Exploration Program.
We acknowledge the VizieR catalogue access tool, provided by CDS, Strasbourg, France.
We acknowledge computational and data storage resources provided by the Tsinghua Astrophysics High-Performance Computing platform at Tsinghua University. 

\vspace{12pt}
{\bf Funding:}

Y.K.J., K.-H.H., S.-J.C., Y.-H.R., S.-M.C., D.-J.K., H.-W.K., S.-L.K., C.-U.L., D.-J.L., Y.L., and B.-G.P. acknowledge support by KASI under the R\&D program (project No. 2024-1-832-00 and 2024-1-832-01) supervised by the Ministry of Science and ICT.
T.S. is supported by JSPS KAKENHI Grant Number 
JP24253004, JP26247023, JP16H06287 and JP22H00153.
%
% Individual acknowledgements
W.Zang, H.Y., S.M., R.K., J.Z., and W.Zhu acknowledge support from the National Natural Science Foundation of China (Grant No. 12133005). W.Zang acknowledges support from the Harvard-Smithsonian Center for Astrophysics through a CfA Fellowship. 
W.Zhu acknowledges science research grants from the China Manned Space Project, grant No.\ CMS-CSST-2021-A11. 
J.C.Y. and I.-G.S. acknowledge support from U.S. NSF Grant No. AST-2108414. 
Y.S. acknowledges support from BSF Grant No. 2020740.
R.P. and J.S. were funded by Polish National Agency for Academic Exchange grant ``Polish Returns 2019.'' 
C.H. was funded by grants from the National Research Foundation of Korea (2019R1A2C2085965 and 2020R1A4A2002885). 
D.P.B., A.B., S.I.S., and G.O. were funded by 
NAA grants 80NSSC20K0886 and 80NSSC24M0022.
\vspace{12pt}
{\bf Competing interests:} 

The authors declare that they have no competing interests.

\vspace{12pt}
{\bf Author contributions:} 

%Weicheng Zang
W.Zang identified the planetary anomaly in OGLE-2016-BLG-0007 and analyzed the event, conducted the frequency analysis, and wrote portions of the manuscript.
%Youn Kil Jung
Y.K.J. performed the sensitivity analysis and wrote portions of the manuscript.
%Jennifer C. Yee
J.C.Y. led the writing of the manuscript and contributed to the analysis.
%Kyu-Ha Hwang
K.-H.H. carried out the systematic search for events in the KMTNet survey and identified this event in the KMTNet data.
%Hongjing Yang
H.Y. re-reduced the KMTNet data and contributed to the analysis.
%Andrew Gould
A.G. is the scientific principal investigator of the KMTNet microlensing survey, contributed to the analysis and provided feedback on the manuscript.
%Shude Mao
S.M. contributed to the analysis and provided comments on the manuscript.
%---------------
%Michael D. Albrow
M.D.A. contributed to the software used in this work and the analysis of data and provided comments on the manuscript.
%Sun-Ju Chung
%Yoon-Hyun Ryu
%In-Gu Shin
%Yossi Shvartzvald
S.-J.C., Y.-H.R., I.-G.S., and Y.S. contributed to the analysis of the KMTNet data and  provided comments on the manuscript.
%Cheongho Han
C.H. conceived of and designed the KMTNet project and contributed to the analysis.
%Chung-Uk Lee
C.-U.L. is the principal investigator of KMTNet and provided comments on the manuscript.
%---------------
%Sang-Mok Cha
%Dong-Jin Kim
%Hyoun-Woo Kim
%Seung-Lee Kim
%Dong-Joo Lee
%Yongseok Lee
%Byeong-Gon Park
%Richard W. Pogge
S.-M.C., D.-J.K., H.-W.K., S.-L.K., D.-J.L., Y.L., B.-G.P., and R.W.P. contributed to KMTNet operations.
%---------------
%Xiangyu Zhang
%Renkun Kuang
%Hanyue Wang
%Jiyuan Zhang 
%Zhecheng Hu
%Wei Zhu
X.Z., R.K., H.W., J.Z., Z.H., and W.Zhu contributed to the analysis.

%OGLE
A.U. is the principal investigator of the OGLE project, and was responsible for the
reduction of the OGLE data. A.U., P.M. J.S., R.P., M.K.S., I.S., P.P.,
Sz.K., K.U., K.A.R., P.I., M.W., and M.G. collected the OGLE
photometric observations, commented on the
results and on the manuscript.

% MOA
T.S. led the MOA project and organized the MOA observations.
I.A.B contributed to the MOA data analysis.
F. A., R. B., A. B., H. F., A. F., R. H., Y. H., S. I. S., Y. I., R. K., N. K., Y. M., S. M., Y. M., G. O., C. R., N. J. R., Y. S., D. S., M. T., P. J. T., A. V., H. Y., and K. Y.
contributed to the MOA observation and commented on the manuscript.

\vspace{12pt}
{\bf Data and materials availability:}

Our observed light curves of OGLE-2016-BLG-0007, the pyDIA color-magnitude diagram, calculated KMTNet sensitivity functions, and a machine-readable version of  our planet sample (Table S3) are available at Dryad \footnote{https://datadryad.org/dataset/doi:10.5061/dryad.ksn02v7cg}.  The Dryad repository also contains codes for the \textsc{AnomalyFinder}, the frequency analysis, and the false alarm probability calculation.
The KMTNet event tables and links to the data are available online at https://kmtnet.kasi.re.kr/ulens/
\cite{KimKim18_EF, Kim18EF}. 
The results of the population model fitting are listed in Table S5. The OGLE-III star catalog \cite{OGLEIII} and limb-darkening coefficient table \cite{Claret11} are available via the VizieR catalogue access tool, CDS, Strasbourg, France.

\clearpage
% Supplement Numbering
\renewcommand{\thefigure}{S\arabic{figure}}
\renewcommand{\thetable}{S\arabic{table}}
\renewcommand{\theequation}{S\arabic{equation}}
\renewcommand{\thesection}{S\Alph{section}}

\setcounter{figure}{0} 
\setcounter{table}{0} 
\setcounter{equation}{0} 

\begin{centering}
\huge{Supplementary Materials for}
\vspace{12pt}

\large{\bf Microlensing events indicate that super-Earth exoplanets are common in Jupiter-like orbits}

\vspace{12pt}
\normalsize{Weicheng Zang*, Youn Kil Jung*, Jennifer C. Yee, Kyu-Ha Hwang,
Hongjing Yang, Andrzej Udalski, Takahiro Sumi, Andrew Gould, Shude Mao*,
Michael D. Albrow, Sun-Ju Chung, Cheongho Han, Yoon-Hyun Ryu, In-Gu Shin, Yossi Shvartzvald, Sang-Mok Cha, Dong-Jin Kim, 
Hyoun-Woo Kim, Seung-Lee Kim, Chung-Uk Lee*, Dong-Joo Lee,
Yongseok Lee, Byeong-Gon Park, Richard W. Pogge,
Xiangyu Zhang,
Renkun Kuang, Hanyue Wang, Jiyuan Zhang, 
Zhecheng Hu, Wei Zhu,
Przemek Mr\'{o}z, Jan Skowron, Rados{\l}aw Poleski, Micha{\l} K. Szyma\'{n}ski, 
Igor Soszy\'{n}ski, Pawe{\l} Pietrukowicz, Szymon Koz{\l}owski, Krzysztof Ulaczyk,
Krzysztof A. Rybicki, Patryk Iwanek, Marcin Wrona, Mariusz Gromadzki,
Fumio Abe, Richard Barry, David P. Bennett, Aparna Bhattacharya,
Ian A. Bond, Hirosane Fujii, Akihiko Fukui, Ryusei Hamada, Yuki Hirao,
Stela Ishitani Silva, Yoshitaka Itow, Rintaro Kirikawa, Naoki Koshimoto,
Yutaka Matsubara, Shota Miyazaki, Yasushi Muraki, Greg Olmschenk,
Cl\'{e}ment Ranc, Nicholas J. Rattenbury, Yuki Satoh, Daisuke Suzuki,
Mio Tomoyoshi,  Paul . J. Tristram, Aikaterini Vandorou,
Hibiki Yama, and Kansuke Yamashita}
\vspace{12pt}

\normalsize{*Corresponding authors:  weicheng.zang@cfa.harvard.edu, ykjung21@kasi.re.kr, leecu@kasi.re.kr,  smao@mail.tsinghua.edu.cn.}
\vspace{12pt}
\end{centering}

\noindent{\bf The PDF file includes:}

Materials and Methods

Supplementary Text

Figs. S1 to S9

Tables S1 to S5

References \textit{(47-127)}

\clearpage

\baselineskip24pt

\section*{Materials and Methods}

\subsection{OGLE-2016-BLG-0007}

\subsubsection{Observations}

OGLE-2016-BLG-0007 was identified as a candidate microlensing event by the Optical Gravitational Lens Experiment (OGLE) Early Warning System \cite{Udalski1994,Udalski2003} on 10 February 2016, 
 at right ascension $17^{\rm h}53^{\rm m}25^{\rm s}.45$ and declination $-29^\circ38'32''.10$ (J2000 equinox), equivalent to
Galactic coordinates longitude $0^\circ.26891$, latitude $-1^\circ.82254$. On 17 March 2016, it was independently identified by the Microlensing Observations in Astrophysics (MOA) collaboration \cite{Bond01} and assigned the designation MOA-2016-BLG-088. The event was also found by the KMTNet post-season EventFinder algorithm \cite{KimKim18_EF, Kim18EF} and designated KMT-2016-BLG-1991. Following standard convention, we adopt the OGLE designation because it was the first survey to identify the event.

The light curve is shown in Figure 2. The OGLE data were taken using the 1.3m Warsaw Telescope with a 1.4 ${\rm deg}^2$ field of view (FoV) camera at the Las Campanas Observatory in Chile \cite{Udalski15_OGLEIV}. The event is located in the OGLE BLG501 field (where ``BLG" indicates the field is toward the Galactic Bulge), which was observed with a cadence of $\sim~1~{\rm hr}^{-1}$. The MOA observations were carried out with the 1.8 m MOA-II Telescope with a 2.2 ${\rm deg}^2$ FoV camera at Mount John Observatory in New Zealand \cite{Sako2008}. The event lies in the MOA gb5 field (where ``gb" indicates the field is toward the Galactic Bulge), which was observed with a cadence of $\sim~4~{\rm hr}^{-1}$. The KMTNet data were taken from three identical 1.6m telescopes equipped with $4~{\rm deg}^2$ FoV cameras located in Australia (referred to as KMTA), Chile (KMTC), and South Africa (KMTS) \cite{Kim16_KMTNet}. The target appears in two slightly offset KMT fields, BLG02 and BLG42 (see fig. 12 of \cite{KimKim18_EF}), with a combined cadence of $\sim~4~{\rm hr}^{-1}$. In Figure 2, the KMTNet data are labeled by site and field, e.g.``KMTC02" corresponds to data from field BLG02 taken by the KMTC telescope. For OGLE and KMTNet, the images were primarily taken in the $I$ band, with 9\% of KMTC data, 5\% of KMTS data, and occasional OGLE data taken in the $V$ band for source color measurements. The MOA images were mainly acquired in the MOA-Red band, which is roughly the sum of the standard Cousins $R$ and $I$ bands. 

Only the $I$ band and MOA data were used in the light curve analysis to determine the best-fitting model. Because microlensing is achromatic, data taken in different bands differ by only a scale factor and offset. The astrophysical effect is identical except for the limb-darkening effect, so converting between different flux systems or observational bandpasses is unnecessary. In the fitting, we scaled the model to the individual datasets and accounted for the stellar limb-darkening effect (which is relatively minor but bandpass dependent), see below. In Figure 2, all data have been scaled to the OGLE flux system. The $V$ band data are lower signal-to-noise, so they were not used in the model fitting and are not shown in Figure 2. They were only used for determining the source color.

The data used in the light-curve analysis were reduced using pipelines based on the difference image analysis (DIA) technique \cite{Tomaney1996,Alard1998}, as implemented by OGLE \cite{Wozniak2000}, MOA \cite{Bond01}, and the \textsc{Tender-Loving-Care (TLC) pySIS} \cite{Albrow09, Yang23_pySIS} pipeline for KMTNet. For the KMTC02 data, we used \textsc{pyDIA} \cite{pyDIA} photometry  to measure the source color. This pipeline provides the $I$ and $V$ band light curves and a catalog of field-star photometry on the same magnitude system. \textsc{pyDIA} $I$-band magnitudes were calibrated using the OGLE-III star catalog \cite{OGLEIII}. 
The 9-yr OGLE data show the source is variable but the variability is long-term ($\sim10~\mathrm{yr}$) and low-level ($\sim 0.03~\mathrm{mag}$). Therefore, for this event we only used the data from 2016, when the microlensing effect occurred. 

\subsubsection{Light-curve analysis}

Because of the bump-type anomaly, a point-lens (1L1S) model is insufficient to fit the light curve of OGLE-2016-BLG-0007. The most likely explanation for the bump is a planet orbiting the lens star, which corresponds to a binary-lens (2L1S) model. An alternative model could be a single star lensing a two-source system (a 1L2S model) \cite{Gaudi98}.

\vspace{12pt}\noindent{\bf S.1.2.1 Analytic estimates}\vspace{12pt}

Excluding the data points around the bump-type anomaly ($7655 < {\rm HJD}^{\prime} < 7659$), we fitted the light curve with a point-source, point-lens (PSPL) model using \textsc{MulensModel} \cite{MulensModel} to determine the parameters of the underlying stellar event, finding:
\begin{equation}
    (t_0, u_0, t_{\rm E}) = (7498.4, 1.253, 73.8~{\rm days}).
\end{equation}
Then, we used a heuristic analysis method \cite{Hwang22AF2,Ryu22_MP1} to estimate the planetary parameters for the 2L1S model: $s$, $q$, and $\alpha$, where $s$ is the separation between the host star and planet as a fraction of the Einstein radius, $q$ is the mass of the planet divided by the mass of the host star, and
$\alpha$ is the angle of the source trajectory with respect to the axis connecting the two lensing bodies. We also estimated $\rho \equiv \theta_* / \theta_{\rm E}$, where $\theta_*$ is the angular radius of the source star.

The time of the anomaly is $t_{\rm bump} = 7656.6$. The offset from the peak, $\tau_{\rm bump}$, and the offset from the host star, $u_{\rm bump}$, both in units of $\theta_{\rm E}$, are
\begin{eqnarray}
  \tau_{\rm bump} &=& \frac{t_{\rm bump}-t_{0}}{t_{\rm E}} = 2.144; \\
  u_{\rm bump} &=& \sqrt{u_{0}^{2}+\tau_{\rm bump}^{2}} = 2.483.
\end{eqnarray}
The planetary caustic is a closed curve in the source plane for which the magnification of a point source would be infinite. It represents the region over which the planet has the largest effect on the source star's light. Because the position of the planetary caustic along the binary axis is $|s - s^{-1}| \simeq u_{\rm bump}$ from the host star \cite{GouldLoeb92}, we obtained
\begin{eqnarray}
& & s_{+} \sim \frac{\sqrt{u_{\rm bump}^2 + 4} + u_{\rm bump}}{2} = 2.83;\\
& & s_{-} \sim \frac{\sqrt{u_{\rm bump}^2 + 4} - u_{\rm bump}}{2} = 0.36;\\
& & \alpha_{+} = 2\pi - \tan^{-1}\frac{u_0}{\tau_{\rm bump}} = 5.75~{\rm rad};\\
& & \alpha_{-} = \pi - \tan^{-1}\frac{u_0}{\tau_{\rm bump}} = 2.61~~{\rm rad};
\end{eqnarray}
where $s_{+}$ and $\alpha_{+}$ correspond to the values for the Wide configuration and $s_{-}$ and $\alpha_{-}$ are for the Close configuration in the heuristic method. For the Wide  configuration, the single planetary caustic lies along the binary axis, so $s = s_{+}$ and $\alpha = \alpha_{+}$. By contrast, for the Close solution, there are two planetary caustics with some vertical separation depending on $q$. The locations of these caustics are described by equations 4 and 14 in \cite{Han06}, which we use in the next section to derive initial starting values for $s$ and $\alpha$ for the formal fitting.

There is an abrupt change in magnification at the onset of the bump-type anomaly, indicating that the source has passed over a caustic. A point source crossing a caustic is always magnified by a factor of at least three \cite{WittMao95}, 
so the lower observed $A_{\rm bump}$ indicates that the 
source star cannot be approximated as a point source. Instead, the angular size of the source affects the observed magnification, referred to as the finite source effect
\cite{MaoPaczynski91}.
For the Wide case, the form of the anomaly suggests the planetary caustic is likely to be smaller than the source. Under these conditions, the source size for this Wide (+) case can be estimated to be
\begin{equation}
    \rho_{+} \geq \frac{\Delta t_{\rm bump}}{2 t_{\rm E}} \sim 0.014,
\end{equation}
where $\Delta t_{\rm bump} \sim 2.1~\mathrm{days}$ is the observed full width of the bump-type anomaly, and we assume this is a lower limit because the caustic crossing time is $\leq2t_{\rm E}\rho$ for a larger source. The excess magnification, $\Delta A$, is then \cite{GouldGaucherel97}
\begin{equation}
    \Delta A = \frac{2q_{+}}{\rho_{+}^{2}}
\end{equation}
where $q_{+}$ is the planet-star mass ratio for the Wide case.
From the light curve, we obtained the peak magnitude of the anomaly, $I_{\rm bump, peak}=15.42$. The PSPL model yielded the baseline magnitude at the time of the anomaly $I_{\rm bump, base}=15.48$ (Fig. 2C) and a source magnitude of $I_{\rm S} = 15.52$. 
The normalized lens-source separation at the anomaly is $u = s - 1/s$. The unperturbed magnification is $A_0 \equiv (u^2 + 2) / u\sqrt{u^2 + 4} = K / \sqrt{K^2 - 4}$ where $K \equiv u^2 +2$. Noting that $K^2 - 4 = (s^2 - 1/s^2)^2$ yields $A_0 = (s^4 + 1)/(s^4 - 1)$. 
Thus, after some algebra,
\begin{equation}
    \Delta A = A_{\rm bump} - \frac{s^4 + 1}{ s^4 - 1} = 0.06,
\end{equation}
where $A_{\rm bump} = 10^{-0.4(I_{\rm bump, peak} - I_{\rm S})}$ and $s = 2.83$.
Combining Equations S9 and S10, we estimated
\begin{equation}
    q_{+} = \frac{\Delta A \rho^{2}}{2} \geq 6.0 \times 10^{-6}. 
\end{equation}

For the Close cases, there are two, triangular planetary caustics whose separation from each other increases as $q^{1/2}$ \cite{Han06}. A large source that envelops both of the triangular caustics tends to generate nearly zero excess magnification \cite{GouldGaucherel97}, so we expect that the source size in the Close cases, $\rho_{-}$, is close to or smaller than the caustic, $\rho_{-} < \rho_{+}$.  In addition, when $q$ is small, the triangular caustics have a strongly demagnified region between them that the source will pass through either before or after crossing the caustic. No such demagnification is apparent in the light curve, which implies that the mass ratio for the Close cases, $q_{-}$, would be large and the triangular caustics would be widely separated,  so, $q_{-} > q_{+}$.

\vspace{12pt}\noindent{\bf S.1.2.2 MCMC analysis}\vspace{12pt}

Setting the initial parameters as the values from the heuristic analysis above, we conducted a numerical analysis using Markov chain Monte Carlo (MCMC) $\chi^2$ minimization using the \texttt{emcee} ensemble sampler \cite{emcee} followed by a downhill algorithm \cite{scipy} to yield the minimum $\chi^2$. We employed the contour integration code  \texttt{VBBinaryLensing} \cite{Bozza2010,Bozza2018} to calculate the magnification as a function of time, $A(t)$. 

The fitting process includes seven free parameters that determine $A(t)$ for a static binary lens ($t_0$, $u_0$, $t_{\rm E}$, $\rho$, $s$, $q$, $\alpha$).
For each data set $i$, we introduced two additional free parameters, $f_{{\rm S},i}$ and $f_{{\rm B},i}$, to represent the source flux and any blended flux, respectively. The blended flux is the fixed component of the measured flux that represents the flux from any other stars, such as the lens, that may be too close to the source to be resolved (and measured) separately. Then, the observed flux $f_{i}(t)$ was modeled as 
\begin{equation}
    f_{i}(t) = f_{{\rm S},i} A(t) + f_{{\rm B},i}.
\end{equation}
Because $f_{{\rm S},i}$ and $f_{{\rm B},i}$ are linear, for each model $A(t)$, we solved for $f_{{\rm S},i}$ and $f_{{\rm B},i}$ directly.

In principle, if the flux parameters, $f_{{\rm S},i}$ and $f_{{\rm B},i}$, are measured on an absolute flux scale, they should always be $\ge 0$. However, small amounts of negative blending are possible in microlensing events because there is a non-uniform background of unresolved dwarf stars. The maximum amount of negative blending is expected to be approximately equal to the flux from a dwarf star.

Allowing the blended flux parameter to vary,
we find the blend flux parameter for the OGLE dataset, $f_{\rm B, OGLE}$, which is measured on an absolute flux system, has a negative value: $f_{\rm B, OGLE} = -0.77 f_{\rm S, OGLE}$ where $f_{\rm S, OGLE}$ is the source flux measured for the OGLE dataset. 
The source is much brighter than a typical dwarf star in this field (we show below that the source is a red-clump giant), so
such a large negative value is unphysical.

To verify that the negative blending does not indicate some other effect we have not accounted for, we compared the $\chi^2$ of the fit
allowing $f_{\rm B, OGLE}$ to be a free parameter to a fit with $f_{\rm B, OGLE} = 0$. This
led to an improvement of only $\Delta\chi^2 = 9$ ($3\sigma$). Fluctuations due to low-level systematics in microlensing light curves with similarly low statistical significance are common, so there is no evidence that the unphysical blending is meaningful. As discussed below, the source star likely accounts for almost all of the baseline flux, so we fixed $f_{\rm B, OGLE} = 0$.

Table S1 lists the parameters of the three 2L1S solutions found by the numerical analysis, and the caustic structures of their best-fitting solutions are shown in Figure S2. For the Close cases, we label the two solutions as Close Upper and Close Lower based on their caustic positions (Figure S2).  The final parameters for both the Wide and Close cases determined by the numerical analysis are consistent with the estimates from the heuristic analysis.

The Wide solution provides the best fit to the observed data in terms of $\chi^2$, while the Close Upper and Close Lower solutions are disfavored by $\Delta\chi^2 =$ 38.5 ($>6\sigma$) and 115.2 ($>10\sigma$), respectively. Figure S3 compares the Wide and Close solutions. The difference in $\chi^2$ arises mostly from the region of the bump anomaly. The Wide model is a better fit to the anomaly except for the KMTS data at ${\rm HJD}^{\prime} \sim 7652.3$. We therefore exclude the Close Upper and Close Lower solutions. 

Due to the source variability and the low magnification of this event, we did not include the microlensing parallax effect in our analysis \cite{Gould1992,Gould2000,Gouldpies2004}. Including lens orbital motion in the model \cite{MB09387, OB09020} did not improve the goodness of fit. 

\vspace{12pt}\noindent{\bf S.1.2.3 1L2S models}\vspace{12pt}

We also investigated fitting binary source, point lens (1L2S) models. For such models, the light curve is the superposition of the single-lens single-source curves of two sources, so the total magnification, $A_{\lambda}(t)$, in the photometric filter $\lambda$ is \cite{MB12486}
\begin{eqnarray}
& & A_{\lambda}(t) = \frac{A_{1}(t)f_{1,\lambda} + A_{2}(t)f_{2,\lambda}}{f_{1,\lambda} + f_{2,\lambda}} = \frac{A_{1}(t) + q_{f,\lambda}A_{2}(t)}{1 + q_{f,\lambda}}, \\
& & q_{f,\lambda} = \frac{f_{2,\lambda}}{f_{1,\lambda}}, 
\end{eqnarray}
where $f_{j,\lambda}$ and $A_{j}(t)$ ($j = 1, 2$) respectively represent the flux in the photometric filter $\lambda$ and magnification of each source. Each magnification curve, $A_{j}(t)$, is defined by three parameters $t_{0,j}$, $u_{0, j}$, and $t_{\rm E}$ where $t_{0, j}$ and $u_{0, j}$ are the same as for a PSPL model but for each source separately. We adopted $q_{f, I}$ as the flux ratio between the secondary and the primary sources for the OGLE and KMTNet data used in the light-curve analysis and $q_{f, {\rm MOA}}$ for MOA.

The best-fitting parameters for the 1L2S model are presented in Table S1. The best-fitting 1L2S model is significantly disfavored by $\Delta\chi^2 = 146~ (12\sigma)$ compared to the best-fitting  2L1S model. If the 1L2S model were correct, the scaled source radius of the second source would be constrained to $\rho_2 = (0.479 \pm 0.034) \times 10^{-2}$; such a second source would have an absolute magnitude of $M_{I,2} \sim 8.6~$mag (see below), yielding an angular source radius of $\theta_{*,2} \sim 0.25~$ microarcseconds and a lens-source relative proper motion of $\mu_{\rm rel} = \theta_{*,2}/\rho_2/t_{\rm E} \sim 0.25~{\rm mas~yr^{-1}}$. Such a low $\mu_{\rm rel}$ has a probability of $\sim 1.8 \times 10^{-3}$ given the kinematics of the model of the Milky Way Galaxy that we adopted  and the $\mu_{\rm rel}$ distributions of observed planetary microlensing events \cite{MASADA,Jung23AF8}.  Therefore, we reject the 1L2S model.

\subsubsection{CMD and source properties}

\vspace{12pt}\noindent{\bf S.1.3.1 Measured properties}\vspace{12pt}

We constructed a color-magnitude diagram (CMD) of OGLE-III field stars within $120''$ centered on the baseline object (Fig. S4).  We fitted the magnitude distribution of the red giant clump in the CMD to find its centroid \cite{Nataf13}. It has a $(V-I)$ color and $I$ magnitude of $(V - I, I)_{\rm cl} = (2.69 \pm 0.01, 16.22 \pm 0.01)$. 
 
The microlensing event was detected by observing a change in the flux of a cataloged star, which we refer to as the baseline object. As discussed above, the flux of this object may be divided into a component from the source and a separate, non-varying component arising from other objects within the point-spread function of the images, including flux from the lens. We measured the properties of the baseline object by matching the KMTC02 \textsc{pyDIA} catalog stars to the OGLE-III catalog \cite{OGLEIII} to calibrate the pyDIA magnitudes. 
 
 The baseline object from the KMTC02 images has a color of $(V - I)_{\rm Base} = 2.99 \pm 0.03$ and a magnitude $I_{\rm Base} = 15.49 \pm 0.03$, which is consistent with a red clump giant star in the bulge. The proper motion of this object in the North (N) and East (E) directions,  \mbox{\boldmath$\mu$}$_{\rm Base}(N, E)$,  in Gaia data release 3 (DR3) \cite{GaiaDR3} was 
\begin{equation}
 \mbox{\boldmath$\mu$}_{\rm Base}(N, E) = (-5.63 \pm 0.07, -3.29 \pm 0.11)~\mathrm{mas~ yr}^{-1},
\end{equation}
with a Renormalised Unit Weight Error (RUWE) parameter of 1.01. This proper motion is typical for a star in the Milky Way's bulge \cite{Clarke19} and the RUWE parameter is low, indicating the measurement is well constrained. 

In the KMTC02 images, the baseline object is located at $(150.27 \pm 0.02, 151.18 \pm 0.02)~$ pixel. We measured the location of the magnified source in the difference images to be $(150.278 \pm 0.002, 151.202 \pm 0.002)~$ pixel. The pixel scale of the KMT images is 0.4 arcsec pix$^{-1}$, so the measured astrometric offset  between the KMTC02 baseline object and the magnified source is $9.5 \pm 8.0~$mas.

The source color $(V-I)_{\rm S}$ can be written in terms of the $V$-band flux, $f_{{\rm S}, V}$, and the $I$-band flux, $f_{{\rm S}, I}$ as
\begin{equation}
(V-I)_{\rm S} = -2.5 \log_{10}(f_{{\rm S}, V}/f_{{\rm S}, I})
\end{equation}
Each KMTC02 $V$-band observation was taken within one minute of a KMTC02 $I$-band observation. The magnification $A(t)$ does not depend on the photometric filter and does not change appreciably on a one minute timescale, which is very short relative to $t_{\rm E}$. Combining Equation S16 with Equation S12 and all pairs of KMTC02 $V$ and $I$ observations, we measured the source color to be $(V - I)_{\rm S} = 3.06 \pm 0.07$. This is
 consistent with the color of the baseline object within $1\sigma$. 
 
\clearpage
\vspace{12pt}\noindent{\bf S.1.3.2 The source as the baseline object}\vspace{12pt}

Because of the slight preference for strong negative blending from the light curve fitting, we fixed the blend flux $f_{\rm B, OGLE} = 0$, effectively assigning all of the observed light to the source. We now consider whether this assumption is consistent with the independent information about the baseline object and source color.
Using the OGLE-III catalog, we estimated the density of stars with $15.4 < I < 15.6$ and $V - I > 2.0$ in this region of the sky is $1.1 \times 10^{-3}$ arcsec$^{-2}$, so the probability that an unrelated star dominates the $I \sim 15.5$ mag light is $\sim 3 \times 10^{-7}$. 
Hypothetically, disregarding the light curve fitting, it would still be possible for most of the flux from the baseline object to come from a companion to the source. 
The source $(V - I)_{\rm S}$ color implies it is either a giant or a low-mass dwarf star. If most of the flux from the baseline object comes from a companion, the source would be dwarf star with an absolute magnitude $M_I > 5.5$ and so, $I_{\rm S} > 21.8$. From the light-curve analysis, we found that a model with $I_{\rm S} > 21.8$ is disfavored by $\Delta\chi^2>490$, so we exclude this possibility.

We therefore conclude that the baseline flux is dominated by light from the source star itself. We ascribe the negative blending found in the model fitting 
to the long-term and low-level source variability.
We therefore adopted the source apparent magnitude $I_{\rm S} = 15.52 \pm 0.01$ from the light-curve analysis with $f_{\rm B, OGLE} = 0$ and the Gaia proper motion as the source proper motion.

\vspace{12pt}\noindent{\bf S.1.3.3 Derived source properties}\vspace{12pt}

We obtained the intrinsic color and de-reddened brightness of the source star \cite{Yoo04_CMDMethod} by locating it on the CMD (Fig. S4). The intrinsic color and de-reddened brightness of the red clump are $(V - I, I)_{\rm cl, 0} = (1.06 \pm 0.03, 14.43 \pm 0.04)$ \cite{Bensby2013, Nataf2013}. The difference between these values and the observed values yields a measurement of the interstellar extinction toward the event direction of $A_I = 1.79 \pm 0.04$ and reddening $E(V - I) = 1.63 \pm 0.03$. The intrinsic color and de-reddened brightness of the source star are therefore $(V - I, I)_{\rm S,0} = (1.43 \pm 0.08, 13.73 \pm 0.04)$. 
This corresponds to a K-type giant star located in the Milky Way's bulge [\cite{Bessell1988}, their table III]. Applying a color/surface-brightness relation for giants \cite{Adams2018}, we estimated an angular source radius $\theta_* = 11.0 \pm 1.0$ microarcseconds. The CMD parameters and $\theta_*$ are summarized in Table S2.

We used a color-temperature relation \cite{Houdashelt2000} to estimate the effective temperature of the source  $T_{\rm eff} \sim 4100~$K. The closest value in a table of linear limb-darkening coefficients [\cite{Claret2011}, table u] is $T_{\rm eff} = 4000$, so we derived linear limb-darkening coefficients $u_I = 0.64$ for the $I$ band and $u_R = 0.76$ for the $R$ band assuming $T_{\rm eff} = 4000$, surface gravity $\log g = 1.5$ (which is appropriate for a giant star), metallicity [Fe/H] = 0.0, microturbulent velocity $\xi=0.0~\mathrm{km~s}^{-1}$, and choosing the results for the ATLAS atmospheric models and Least-Square method. For the MOA data, we assumed $u_{\rm MOA} = (u_I + u_R)/2 = 0.70$. Because the coverage of the bump anomaly is sparse,
minor variations in the limb-darkening law or coefficients due to differences in the assumptions or model choices would have almost no effect on the results.

\subsubsection{Lens properties}

Combining the angular source radius, $\theta_*$, and the scaled source radius, $\rho$, we obtained the angular Einstein radius: $\theta_{\rm E} = \theta_*/\rho  = 0.73 \pm 0.10~ \mathrm{mas}$. We then derived the lens-source relative proper motion $\mu_{\rm rel} = \theta_{\rm E}/t_{\rm E} = 3.61 \pm 0.51~ \mathrm{mas~ yr}^{-1}$ (Table S2).

The mass of the lensing object, $M_{\rm L}$, and the lens distance, $D_{\rm L}$, are related to $\theta_{\rm E}$ and the microlensing parallax, $\pi_{\rm E}$, by \cite{Gould1992, Gould2000}
\begin{equation}
    M_{\rm L} = \frac{\theta_{\rm E}}{{\kappa}\pi_{\rm E}};\qquad D_{\rm L} = \frac{\mathrm{1~au}}{\pi_{\rm E}\theta_{\rm E} + \pi_{\rm S}},
\end{equation}
where $\kappa \equiv 4G/(c^2) \mathrm{~au^{-1}} = 8.144$ mas$~M_{\rm Sun}^{-1}$, where au is the astronomical unit, and $\pi_{\rm S}$ is the source parallax. From those properties we derived the planetary mass, $m_{\rm p}$, and the projected host–planet separation, $a_{\bot}$,
\begin{equation}
    m_{\rm p} = q M_{\rm host} ; \qquad a_{\bot} = sD_{\rm L}\theta_{\rm E}
\end{equation} 
where $M_{\rm host} =  M_{\rm L} / (1 + q)$ is the mass of the host star.
For OGLE-2016-BLG-0007, $\theta_{\rm E}$ was measured but $\pi_{\rm E}$ was not, so we could not directly solve Equations S17 and S18 for the physical properties of the lens or planet. We therefore estimated them statistically using a Bayesian analysis based on a Galactic model. We adopted an existing Galactic model \cite{Zang21AF1} but updated it with the source proper motion measurements (Equation S15) from Gaia DR3 \cite{GaiaDR3} . 

We simulated a sample of $10^8$ events from the Galactic model. For each simulated event $k$ with timescale $t_{\rm E,k}$, lens-source relative proper motion $\mu_{{\rm rel},k}$, and Einstein radius $\theta_{{\rm E},k}$, we applied weights
\begin{equation}
    w_{k} = \Gamma_{k}\times p(t_{\rm E,k}) p(\theta_{{\rm E},k}),
\end{equation}
where $\Gamma_{k} = \theta_{{\rm E},k} \times \mu_{{\rm rel},k}$ is the microlensing event rate, and $p(t_{\rm E,k})$ and $p(\theta_{{\rm E},k})$ are the likelihood of $t_{\rm E,k}$ and $\theta_{{\rm E},k}$ given the uncertainties ($\sigma_{t_{\rm E}}$ and $\sigma_{\theta_{\rm E}}$) on $t_{\rm E}$ and $\theta_{\rm E}$  from the model fitting (Tables S1 and S2),
\begin{eqnarray}
& & p(t_{\rm E,k}) = \frac{{\rm exp}[-(t_{{\rm E},k} - t_{{\rm E}})^2/2\sigma^2_{t_{{\rm E}}}]}{\sqrt{2\pi}\sigma_{t_{\rm E}}}, \\
& & p(\theta_{{\rm E},k}) = \frac{{\rm exp}[-(\theta_{{\rm E},k} - \theta_{\rm E})^2/2\sigma^2_{\theta_{\rm E}}]}{\sqrt{2\pi}\sigma_{\theta_{\rm E}}}.
\end{eqnarray}

The resulting posterior probability distributions for $m_{\rm p}$, $M_{\rm host}$, $D_{\rm L}$, and $a_{\bot}$ are shown in Figure S5 and were used to determine the median and 68\% confidence ranges listed in Table 1. The lens (host) star is consistent with being a low-mass dwarf star in the disk of the Milky Way. This would contribute negligible light to the event, relative to the red clump giant source star, consistent with our inferences above.
Assuming a circular orbit and a random orientation of the planet around the host star, we also calculated statistical probability distributions for the semi-major axis, $a$, and the planet orbital period, $P$ (Table 1). 

The median value of the orbital period is $P\sim 40~\mathrm{yr}$, and the $1\sigma$ confidence interval is $\sim30$ to $60~~\mathrm{yr}$. In the model-fitting, we did not detect the orbital motion effect of the planet. This is consistent with the long duration of the orbital period relative to both relevant timescales in the event: the Einstein timescale, $t_{\rm E} = 74~\mathrm{days}$ (or $0.20~\mathrm{yrs}$), and the $0.43~\mathrm{yr}$ between the the peak of the stellar event, $t_0 = 7498.44 $, and the time of the planetary perturbation, $t_{\rm bump}= 7656.5$.

\subsection{Wider microlensing planet sample selection}%\label{sec:meth:sample}}

\subsubsection{Overview}

We combined OGLE-2016-BLG-0007 with previously published \textsc{AnomalyFinder} detections to assemble a statistical sample of planets from the 2016--2019 KMTNet microlensing seasons. 
In those publications, 
the KMTNet images were reduced using a pipeline version of pySIS \cite{Albrow09} to produce the initial photometry, which is publicly available on the KMTNet website \cite{KimKim18_EF,Kim18EF}.
The  \textsc{AnomalyFinder} algorithm was applied to these public data to identify candidate planets.
Some anomalies were already known to be planets or binary stars from previous publications. 
Some anomalies due to binary-star systems were eliminated by visual inspection. 
The remaining anomalies were fitted to determine their mass ratios, $q$. 
If there was a viable solution with $q < 0.06$, the images were re-reduced to produce optimized photometry. For the new planet candidates, these re-reductions were done using a revised TLC version of \textsc{pySIS} that uses more advanced image metrics  \cite{Yang23_pySIS}.
The revised photometry for each event was fitted with a microlensing model, then
 planets and planet candidates were defined as events with solutions that have $q < 0.03$. This is the same procedure and software used identify and analyze OGLE-2016-BLG-0007Lb.

We started by using 2018 and 2019 as representative seasons \cite{Zang22AF4, Gould22AF5, Jung22AF6, Jung23AF8}. 
To increase the number of small planets, we extended the sample to include $\log q < -4$ planet detections from 2016 and 2017 \cite{Hwang22AF2,Zang23AF7}. 

For the statistical sample, we required the planets to have well constrained mass-ratios. Specifically, for solutions within $\Delta\chi^2 < 10$ of the best-fitting solution, we required the difference in the logarithm of the mass ratio between that solution and the best-fitting solution to be small: $\Delta \log q < 0.25$. We also required the remaining solutions to have small uncertainties in $\log q$: $\sigma_{\log q} < 0.2$. Finally, we rejected planets in binary star systems, as well as events with plausible binary-source (non-planetary) solutions. We discuss the impact of each of these cuts below.

Our final sample consists of 63 planets, including 20 planets with $\log{q} < -4$ and 43 planets with $\log{q} > -4$. Table S3 lists the event name, $\log q$ and $s$ for degenerate solutions with $\Delta\chi^2 < 10$ relative to the best-fitting solution, and references for the 63 planets. 

\subsubsection{Detailed sample selection}

In the sample of planets from the 2018--2019 KMTNet seasons and 2016--2017 planets with $\log{q} < -4$, there were a total of 88 planets detected by the \textsc{AnomalyFinder} algorithm. We considered six different factors when evaluating these planets for inclusion in our statistical sample.

First, we considered the 44 anomalous events that were analyzed based on photometry produced prior to the development and widespread application of the revised \textsc{TLC} version of \textsc{pySIS}  \cite{Yang23_pySIS}. The previously analyzed photometry for some of these events was produced using the original \textsc{pySIS} software \cite{Albrow09}, but with the parameters and reference images tuned to the specific event to produce improved photometry relative to the pipeline version of the data that is publicly available. Previous photometry for other events was produced using an earlier version of the revised \textsc{TLC pySIS} software, which did not include all of the updates introduced in \cite{Yang23_pySIS}. 

We found three events with non-negligible differences in the photometry when reduced with the revised \textsc{TLC} version of \textsc{pySIS}.
They are OGLE-2018-BLG-0532, KMT-2018-BLG-1025, and KMT-2019-BLG-1953. 
% OB180532
For OGLE-2018-BLG-0532, a $2\sigma$ tension between the light-curve-based solution and the constraints from blended light was previously reported \cite{OB180532}. We obtained the same tension with the revised pipeline, so for simplicity
we adopted the previously-published results. 
% KB181025
For KMT-2018-BLG-1025 previous work \cite{KB181025} found that the light curve was best fitted by a solution with $\log{q} = -4.081 \pm 0.141$ but there are two degenerate solutions with $(\Delta\log{q}, \Delta\chi^2) = (0.28, 8.4)$ and (0.29, 11.8), respectively. Using photometry from the revised \textsc{TLC} version of \textsc{pySIS} \cite{Yang23_pySIS}, the two degenerate solutions are disfavored by $\Delta\chi^2 > 30$, and the best-fitting solution has $\log{q} = -4.203 \pm 0.137$; we adopted these values. 
% KB191953
For KMT-2019-BLG-1953, the previous report \cite{KB191953} identified a weak anomaly in the high-magnification peak (maximum magnification $A_{\rm max} \sim 900$), which was interpreted as a planet with $\log{q} = -2.706 \pm 0.139$. We found that the eight most magnified data points over the microlensing peak are affected by saturation. After excluding those eight data points, the anomaly is too weak to pass our selection criteria, so we excluded this planet from our sample.

Second, we excluded four planets in binary-star systems (OGLE-2018-BLG-1700 \cite{OB181700}, KMT-2019-BLG-1715 \cite{KB191715}, OGLE-2019-BLG-1470 \cite{OB191470}, and KMT-2019-BLG-1470 \cite{Zang22AF4}),
because the \textsc{AnomalyFinder} algorithm cannot yield a homogeneous sample for planets in binary-star systems. We also excluded known binaries from the planet sensitivity calculation (see below) for consistency.
Including planets around binaries could also introduce
ambiguities to the planet population. For example, three of the four planets could be either circumbinary planets (P-type orbit) or circumstellar planets (S-type orbit) with different implications for the formation and evolution of the planets. 
In all cases, the presence of the stellar companions could influence the formation or evolution of the planets, which is also grounds to exclude them from the 
main sample.

Third, we excluded five planet candidates [KMT-2016-BLG-0212 \cite{Hwang18_0212}, KMT-2018-BLG-2164\cite{Gould22AF5},  KMT-2018-BLG-2718\cite{Gould22AF5}, OGLE-2018-BLG-1554\cite{Gould22AF5}, OGLE-2019-BLG-0344\cite{Jung23AF8}] that have degenerate binary-star solutions ($\log q \geq -1.5$). 
We considered all solutions within $\Delta\chi^2<10$ of the best-fitting model to be viable solutions.
We therefore excluded four planet candidates because they have degenerate binary-star solutions within $\Delta\chi^2<10$ of the planetary solution.
In addition, KMT-2018-BLG-2718 has a binary-star solution with $\Delta\chi^2 = 12.7$. During the anomaly in that event many data points had high sky background due proximity to the moon, so could have low-level systematics, 
so we also excluded it.

Fourth, 
we excluded nine planet candidates because there is a degeneracy \cite{Gaudi98} between the best-fitting 2L1S and 1L2S models:
KMT-2016-BLG-0625 \cite{Shin23_AF9}, KMT-2016-BLG-1105 \cite{Zang23AF7}, OGLE-2018-BLG-1554 \cite{Gould22AF5}, KMT-2018-BLG-0173 \cite{Jung22AF6}, KMT-2019-BLG-0304 \cite{Jung23AF8}, and KMT-2019-BLG-0414 \cite{Jung23AF8}, KMT-2018-BLG-1497 \cite{Jung22AF6}, KMT-2018-BLG-1714 \cite{Jung22AF6}, KMT-2018-BLG-2004 \cite{Gould22AF5}. 
We excluded planet candidates with a $\chi^2$ for the best-fitting 1L2S model, $\chi_{\rm 1L2S}^2$, with 
$\chi^2_{\rm 1L2S} - \chi^2_{\rm 2L1S} \leq 15$, where $\chi_{\rm 2L1S}^2$ is the $\chi^2$ for the best-fitting planetary model. This accounts for eight of the candidates.
In addition to the $\chi^2$ information from the light-curve analysis, we also considered the probabilities from the ``color argument'' for the second source \cite{Gaudi1998} and the ``kinematic argument'' for the lens-source relative proper motion $\mu_{\rm rel}$ \cite{MASADA,Jung23AF8}.
KMT-2018-BLG-0173 has $\chi^2_{\rm 1L2S} -\chi^2_{\rm 2L1S} = 16.3$, but we still excluded it 
for the same reasons of moon proximity as above.

Fifth, we excluded two planets for which any of the degenerate solutions has an uncertainty in $\log q$ of $\sigma_{\log q} \geq 0.20$. This criterion ensured each solution in our sample has a well-constrained mass ratio for the frequency calculation. The two planets were in events KMT-2018-BLG-1988 \cite{KB181988} and OGLE-2018-BLG-1126 \cite{Gould22AF5}. For OGLE-2018-BLG-1126, 
the anomaly was also covered by MOA, so we re-analyzed the event including the MOA data, but the resulting $\sigma_{\log q}$ for both of the degenerate solutions was still $>0.20$. 

Sixth, 
we excluded four planets for which 
the $\log q$ difference, $\Delta\log q$, between any pair of degenerate solutions was
$\Delta\log q > 0.25$. 
This criterion means that the 1-$\sigma$ uncertainty of the combined probability distribution of $\log q$ for all solutions is $<0.20$ as long as the uncertainties in individual solutions, $\sigma_{\log q}$, are $\leq 0.15$.

For the remaining 63 planets, 60 have $\sigma_{\log q} \leq 0.15$ for all solutions while three have at least one solution with $0.15 < \sigma_{\log q} < 0.18$. 
For all three planets with solutions that have $0.15 <\sigma_{\log q} < 0.18$ (OGLE-2017-BLG-1691, \cite{OB171691}; OGLE-2017-BLG-1806, \cite{Zang23AF7}; KMT-2017-BLG-1003, \cite{Zang23AF7}), the maximum $\Delta\log{q}$ $\leq 0.14$, so the uncertainty in $\log q$ from the combined probability distribution for all solutions is $< 0.2$.
For the $\Delta\chi^2$ criteria of degenerate solutions, if we were to adopt a criterion of 16 (equivalent to $4\sigma$) instead of 10, the only excluded planet would be the planet in OGLE-2017-BLG-1806, and if we were to adopt a criterion of 25 ($5\sigma$), only the planets in OGLE-2018-BLG-0596 \cite{OB180596} and KMT-2019-BLG-1216 \cite{Jung23AF8} would be further rejected. 
Therefore, most of the planets ($>95\%$) in our sample have a well-constrained mass ratio regardless of the $\Delta\chi^2$ criteria of degenerate solutions, especially the six planets with $\log{q} < -4.4$. 

Our final sample of 63 planets found in 60 events is listed in Table S3.

\subsection{Sensitivity calculation}

We selected the sample of events for determining the planet sensitivity function, $\mathcal{S}(\log s, \log q)$, from the 2018 and 2019 KMTNet seasons. In the 2018 and 2019 seasons, the KMTNet \textsc{AlertFinder} \cite{Kim18_AF} and \textsc{EventFinder} \cite{KimKim18_EF, Kim18EF} pipelines and subsequent vetting processes identified 2781 and 3303 events, respectively. 

We fitted all non-binary events with a PSPL microlensing model. We tested whether finite-source effects needed to be included in the light curve model by fitting a finite-source--point-lens (FSPL) model to the data. If there was a $\Delta\chi^2$ improvement of $>30$ relative to a point-source model, $\rho$ was considered to be measured, and we adopted the FSPL model as the best-fitting point-lens model. Otherwise, we used the PSPL model. We additionally checked the microlens parallax. If the best-fitting point-lens model including parallax improved the $\chi^2$ by $\Delta\chi^2>30$ relative to a model with zero parallax, the microlens parallax is constrained to be non-zero.  We then used the parallax models to evaluate the criteria listed in Table~S4.

We applied the set of cuts given in Table~S4 using the parameters of the best-fitting point-lens model. 
Cut-1 was used to exclude variable stars with low-amplitude pulsations. Cuts-2 and -3 were used to reject non-microlensing events. Cut-4 ensured that there were a sufficient number of data points, $N_{\rm data}$, in the time interval $t_0 \pm t_{\rm E}$ to detect a planetary signal; $N_{\rm data}$ was evaluated in two halves, one for the rising side of the light curve, $N_{\rm r}$, and one for the falling side, $N_{\rm d}$. Cut-5 eliminated microlensing events in which the peak was located either outside of the observing season or  too close to the beginning or end of the season; $N_{t_{\rm eff}}$ is the number of data points in the range $t_0 \pm t_{\rm eff}$ (where $t_{\rm eff} \equiv u_0\times t_{\rm E}$). Cut-6 was applied because in all planetary events, the source is brighter than 23 mag in the $I$-band. 

Our final sample of events for determining the planet sensitivity consists of 1780 events from 2018 and 1877 events from 2019 (which includes the 31 and 22 events with planets from 2018 and 2019, respectively).

We calculated the sensitivity, $\mathcal{S}_\ell (\log s, \log q)$, for each individual event $\ell$ by injecting
simulated planets with properties $\log s$, $\log q$, and $\alpha$ into the light curve data.
We then evaluated $\mathcal{S}_\ell (\log s, \log q)$ as the fraction of angles, $\alpha$, for a given ($\log s$, $\log q$) pair that produced an anomaly that was recovered by the \textsc{AnomalyFinder} algorithm \cite{Zang21AF1,Zang22AF4}. 
The injected planets were drawn from a uniform grid of $-1 \le \log s \le 1$, $-7.0 \le \log q \le -1.4$, and $0~\mathrm{deg}~\le~\alpha~\le~358~\mathrm{deg}$
with steps of $\Delta \log s = 0.04$, $\Delta \log q = 0.1$, and $\Delta \alpha = 2~\mathrm{deg}$. 
Given a particular set of planet parameters $(\log s, \log q, \alpha)$, we generated a planetary model using the best-fitting values for $t_0$, $u_0$, and $t_{\rm E}$; we also included the two (North and East) components of the microlens parallax vector, $(\pi_{\rm E, N}, \pi_{\rm E,E})$, if the parallax was constrained to be non-zero. 
To generate a planetary model, we also needed a value of the source size, $\rho$. When this was not measured from the original light curve, we estimated it from the source magnitude assuming that the source is in the Galactic bulge following the same procedure as a previous study \cite{Gaudi02}. We estimated the angular source radius by combining the observed source magnitude and color with a 10 gigayear stellar isochrone \cite{Spada17}, and we estimated the lens-source relative proper motion from a model of the Milky Way Galaxy \cite{Jung21}.
Next, we added the planetary model to the residuals from the best-fitting model to the data to generate a simulated light curve, and then ran  \textsc{AnomalyFinder} on the simulated light curve. 

If the \textsc{AnomalyFinder} algorithm recovered the planetary signal from these simulated data, we conducted a series of additional tests. First, we defined the anomaly region as the time range $(t_{\rm{AF}, 0, \rm{anom}}- 3t_{\rm{AF}, \rm eff} <t_{\rm{AF}, 0, \rm{anom}} < t_{\rm{AF}, 0, \rm{anom}} + 3 t_{\rm{AF}, \rm eff})$, where $t_{\rm{AF}, 0, \rm{anom}}$ and $t_{\rm{AF}, \rm eff}$ are the anomaly time and duration parameters reported by \textsc{AnomalyFinder}  \cite{Zang21AF1,Zang22AF4}. 
In this region, we defined several parameters and criteria to eliminate planetary signals that pass the basic \textsc{AnomalyFinder} threshold but would not make it into our final sample. A major source of false positives in this injection/recovery analysis arises from weak, long-timescale anomalies that cannot be distinguished from systematics in the real data. The following criteria ensured that the signal from the planet is concentrated enough in time and strong enough for it to be identified by eye as a real anomaly given the typical level of systematics in the KMTNet data. 

We calculated the total $\chi^2$ contribution from datapoints during the anomaly region defined above  for the best-fitting PSPL model, FSPL model, and planetary (2L1S) model fitted to the simulated data: $\chi^2_{\rm anom, PSPL}$, $\chi^2_{\rm anom, FSPL}$, and $\chi^2_{\rm anom, 2L1S}$. We defined $\Delta A_3$ and $\Delta A_4$ as follows: For a given point at time $t$, $\Delta A (t) = A_{\rm 2L1S}(t) - A_{\rm FSPL}(t)$, where $A_{\rm 2L1S}(t)$ is the magnification of the 2L1S model at time $t$ and $A_{\rm FSPL}(t)$ is the magnification of the FSPL model. Then, $\Delta A_3 = \max( |\Delta A(t+1) - \Delta A(t)|)$ and $\Delta A_4 = \max(|\Delta A(t) / A_{\rm FSPL}(t) |)$. 
We required
 \begin{enumerate}
 	\item{$\chi^2_{\rm anom, PSPL} - \chi^2_{\rm anom, 2L1S} > 50$.} 
	\item{$\chi^2_{\rm anom, FSPL} - \chi^2_{\rm anom, 2L1S} > 30$.}
	\item{the minimum absolute change in magnification due to the planet to be $\Delta A_3 > 0.06$. }
	\item{the minimum fractional magnification due to the planet to be $\Delta A_4 > 0.015$.}
\end{enumerate}

Additionally, before a simulated planet can be considered ``detected", we checked point-lens--binary-source (1L2S) models. We required that 
\begin{enumerate}
	\setcounter{enumi}{4}
	\item{$\chi^2_{\rm 1L2S} - \chi^2_{\rm 2L1S} > 15$ as calculated from the whole light curve. }
\end{enumerate}
This eliminated anomalies that would be considered too ambiguous to be included in the planet sample defined in Section S.2. As a sanity check, we also explicitly checked the $\log q < -4$ planet detections in our sample, because such planets would tend to have the weakest signals, and confirmed that they pass these cuts.

To determine the sensitivity function, $\mathcal{S} (\log s, \log q)$, we summed over the individual event sensitivities $\mathcal{S}_\ell (\log s, \log q)$ for all events in the sample. 
Finally, we integrated over $\log s$, to obtain the sensitivity as a function of $\log q$, $S(\log q)$. We checked for variations in the sensitivity function by calculating it separately for the 2018 and 2019 seasons and for both seasons together. The results are shown in Figure S1 and are nearly identical. 

%%%%%%
\subsection{Frequency calculation}

\subsubsection{Fitted functions}

We followed existing methodology \cite{Alcock1996,mufun,Suzuki16} to determine the mass-ratio function, $f(q)$, by maximizing the likelihood, $\mathcal{L}$:\begin{equation}
 \mathcal{L} = e^{-N_{\rm exp}} \prod_{m=1}^{N_{\rm obs}} f(q_m) \times S(\log{q_m}).
\end{equation}
$N_{\rm obs}$ is the number of planets detected, $q_m$ is the mass ratio for each of the planets detected, and 
$N_{\rm exp}$ is the expected number of planets detected for a given mass-ratio function:
\begin{equation}
  N_{\rm exp} = \int_{10^{-6.0}}^{10^{-1.5}} dq~f(q) \times S(\log{q}).
 \end{equation}
For each planet $m$, we adopted  
\begin{equation}
    \log{q_m} = \sum_{n=1}^{N_{\rm deg}} \log{q_{m,n}}/N_{\rm deg},
\end{equation}
where $\log{q_{m,n}}$ is the value for each degenerate solution $n$, and $N_{\rm deg}$ is the total number of degenerate solutions.

As discussed in the main text, we fitted the data with two fiducial mass-ratio functions. They are
\begin{eqnarray}
&&\mathrm{Power~Law}: \nonumber \\
&&\quad f(q) = \mathcal{A} \biggl(\frac{q}{10^{-4}} \biggl)^{\gamma} 
\end{eqnarray}
where $\mathcal{A}$ is the normalization and $\gamma$ is the power-law index; and
\begin{eqnarray}
&& \mathrm{Double~Gaussian:} \nonumber\\
&&\quad f(q) = \mathcal{A}_1\times10^{n_1(\log{q}-\log{q_{\rm peak,1}})^2} + \mathcal{A}_2\times10^{n_2(\log{q}-\log{q_{\rm peak,2}})^2} .
\end{eqnarray}
The power-law was chosen because it is commonly used for describing astronomical mass functions. It is also the simplest function that differs from a completely flat distribution (i.e., requires the fewest free parameters). The double-Gaussian was chosen to produce a bi-modal distribution and because it is the simplest expression for describing two distinct, non-infinite populations.
We adopted uniform priors for all parameters and set bounds on 
the peaks for the two Gaussians:$-6.0 < \log{q_{\rm peak, 1}} < -3.5$ and $-3.0 < \log{q_{\rm peak, 2}} < -2.0$. We obtained the posterior probability distributions of the parameters using MCMC analysis. The resulting parameters are presented in Table S5, including the best-fitting values, the median values, and $68\%$ confidence interval for each parameter. The best-fitting model with the double-Gaussian function is favored over the power-law model by 
$-2\Delta\ln {\cal L} = 22.6$.
Figure 3 shows these functions with 1$\sigma$ confidence intervals constructed from the MCMC analysis as the region within $-2\Delta\ln {\cal L} =1$ of the best-fitting models.

Poisson statistics determine the uncertainties for calculating the preference for the double-Gaussian over the power-law but are difficult to infer from Figure 3 because of the logarithmic scale. That figure is also shows errors on the both the data and the models even though, formally, there are no errors on the data. Figure S6 is an alternative format for visualizing the results. It compares the observed number of planets in various $\log q$ bins, $n_{\rm obs}$, to the number of planets, $n_{\rm exp}$, predicted by each model to be observed given our data and sensitivity function. We use the approximation $\sigma = \sqrt{n_{\rm exp}}$ to illustrate the uncertainties on these predicted values.

The preference for the double-Gaussian over the power-law comes in large part from under-abundances of observed planets in three regions of the $\log q$ distribution: $\log q = [-6.0, -5.0]$, $[-3.6, 3.0]$, and $[-2.0, -1.5]$. These regions were determined from inspection of the cumulative distribution in Figure 3 and differ slightly from the bin widths in Figure  3A.
Figure S7 shows the Poisson distributions for the expected number of planets in each of these bins compared to the observed number of planets (1, 3, and 3 for the $\log q = [-6.0, -5.0]$, $[-3.6, 3.0]$, and $[-2.0, -1.5]$ bins, respectively). These distributions are a more accurate representation of the uncertainties than the $\sqrt{n_{\rm exp}}$ approximation used above. They show the power-law model has a low probability for observing such a small number of planets in these bins. 
Figure S7 shows the probability, ${\cal P}$,
in each of the six panels corresponding to the total magenta area in each panel, i.e.,
\begin{equation}
 {\cal P} = e^{-n_{\rm exp}}\sum_{\nu=0}^{n_{\rm obs}} {n_{\rm exp}^\nu\over \nu!}
\end{equation}
where $\nu$ is a dummy variable, $n_{\rm obs}$ is the number of observed planets 
in the bin (in contrast to $N_{\rm obs}$ the total number of planets in the sample), and $n_{\rm exp}$ is the number expected for that bin 
from the model.

We also investigated fitting a broken power-law model \cite{Suzuki16}, which has four free parameters.
Our data disfavors the previous model [(5), the ``$q_{\rm br}$-Fixed, MOA-only" model in their table 5]
by $-2\Delta\ln {\cal L} > 14.8$ relative to the single-power-law model, even when we allowed the parameters to vary within the quoted $1\sigma$ ranges. Allowing all parameters to be free-parameters, 
we found a broken-power-law function is favored over a power-law function by $-2\Delta\ln {\cal L} = 5.5$ (i.e., it has a false-alarm probability of 6.4\%, so the improvement is not significant). The best-fitting broken power law model has a break at $\log{q_{\rm break}} \sim -5.0$ and is shown in Figure S8. For this solution, $\gamma = -0.585$ at $\log q > \log{q_{\rm break}}$, 
which is the same as the single-power-law index within its uncertainties (Table S5). For $\log q < \log{q_{\rm break}}$, the power-law index is $> 0$ but is poorly constrained because 
there is only one detected planet with $\log q$ below $\log{q_{\rm break}}$.
We also tested a broken-power-law model with $\log{q_{\rm break}} \equiv -3.77$ [as in previous work \cite{Suzuki16}] and found the goodness of fit improves by $-2\Delta\ln {\cal L} =1.7$, which is also not significant. Hence, although the distribution appears to be flat for $\log q < -4.25$, we reject the broken-power-law model as unsupported by the mass-ratio distribution.

\subsubsection{False-alarm probability}

The double-Gaussian model has more free parameters than the single-power law, so  $-2\Delta\ln {\cal L}$ is not sufficient for model comparison.
However, as we argued in the main text,
either runaway gas accretion or gravitational instability could plausibly produce a bimodal planet distribution, which might produce a distinct population of massive, gaseous planets (similar to Jupiter). In core-accretion theory, these two populations could correspond to planets whose growth terminated in Phase II versus Phase III of planet formation, i.e., before or after the onset of runaway gas accretion \cite{Pollack96}.
We therefore regard a two-population model
as physically motivated whereas the single power-law
model is not. 
There is no simple numerical factor that can describe this prior expectation in a Bayesian sense.
For our analysis, we chose a  6-parameter, double-Gaussian model to represent the two populations.

For model comparison, we considered the false alarm probability (FAP). 
Under the assumption that the power-law model is correct, 
we calculated the probability that, solely due to statistical noise,
the data would appear to be more consistent with a double-Gaussian
model with $-2\Delta\ln {\cal L} > 22.6$ (the observed improvement of the double-Gaussian over the single power law).

In the limit of large-number statistics, $-2\Delta\ln {\cal L} \rightarrow \Delta\chi^2$ for which the false-alarm probability would be given by
$p= (1 + \Delta\chi^2/2)\exp(-\Delta\chi^2/2) = 1.49 \times 10^{-4}$.
However, because the number of detections is subject to Poisson statistics rather than the Gaussian limit, we expect that
$p$ will be greater than this. Therefore, we carried out an explicit calculation of the FAP as follows:
\begin{enumerate}
	\item{Assume an underlying $q_{\rm observed}$ function equal
to the best-fitting power-law model, and multiply it by the sensitivity function, taking into account the two times higher sensitivity for $\log q<-4$ due
to 4 seasons of observations rather than 2.}
	\item{Generate a million simulated data sets each containing 63 planets.}
	\item{Fit each of these simulated populations with both double-Gaussian and single-power law models.}
	\item{Calculate the fraction with $-2\Delta\ln {\cal L} > 22.6$.}
\end{enumerate}
Out of 1 million trials, 160 had $-2\Delta\ln {\cal L} > 22.6$.  Thus
the measured FAP is
$p = (160  \pm \sqrt{160}) \times 10^{-6} = (1.60 \pm 0.13) \times10^{-4}$,
which is very similar to but slightly higher than the value calculated assuming Gaussian statistics.

{\subsubsection{Potential correction factor for excluded planets}

Twenty-five planets or planet candidates were eliminated from our original sample through the cuts described above. The effect of these cuts on our inferences about the underlying planet population primarily depends on whether or not our calculation of the sensitivity reproduced those cuts. In our sensitivity calculation (described above), we explicitly accounted for potential degeneracies between 2L1S and 1L2S models. Hence, the elimination of nine planet candidates due to this degeneracy and the two candidates that also have degenerate binary solutions was already accounted for in our calculation. In addition, we excluded events with clear binary signals from the calculation of the planet sensitivity, which accounted for the four planets eliminated for being in binary star systems. One planet was eliminated for failing the \textsc{AnomalyFinder} cuts. This left six planets whose mass ratios were too uncertain to include in our mass-ratio distribution and three planet candidates that could instead be binary stars. Our sensitivity calculation did not evaluate potential ambiguities or uncertainties in the $\log q$ of the injected planetary signals. This gives a minimum correction factor to the overall frequency of planets of 9.5\% to account for the known planets and a maximum of 14.3\% if all of the objects with binary solutions are real planets.

Assuming a uniform distribution over this interval and
Poisson statistics,
leads to a nominal correction factor calculated from the mean
of $11.9\% \pm 4.6\%$. 
This additional (4.6\%)
uncertainty is small compared to the $\sqrt{1/63}=12.6\%$
Poisson uncertainty in the normalization derived above, and if included, would 
only increase the overall uncertainty from 12.6\% to 13.4\%.
Because of the large uncertainties in this correction factor and the minimal effect on the results relative to the existing uncertainties, we do not include it in our reported values for the planet frequency.

\subsubsection{Checking effect of $\log q$ uncertainties}

We also considered
the uncertainties in $\log q$ for the KMTNet planets.
For our sample of 63 planets, we assumed that degenerate solutions were equally probable. To marginalize over the $\log q_j$
uncertainties, we drew $\mathcal{K}$ samples 
 for the $n$th solution of planet $m$ assuming a 
Gaussian distribution. 
Equations S22 and S23 become 
\begin{eqnarray}
 &   \mathcal{L} = e^{-N_{\rm exp}} \prod_{m=1}^{N_{\rm obs}} \biggl(\frac{1}{N_{\rm deg}}\sum_{n=1}^{N_{\rm deg}}\frac{1}{\mathcal{K}}\sum_{z=1}^{\mathcal{K}}f(q_{mnz}) \times S(\log{q_{mnz}})\biggl),\\
 & N_{\rm exp} = \int_{10^{-6.0}}^{10^{-1.5}} dq~f(q) \times S(\log{q})
\end{eqnarray}
and the models were re-fitted to the data.
We found that the best-fitting parameters change by $<0.1\sigma$ for both the power-law function and the double-Gaussian function. 

When we included the uncertainty of $\log{q_j}$ and weighted the different solutions by $e^{-\Delta\chi^2/2}$, where $\Delta\chi^2$ is the $\chi^2$ difference compared to the best-fitting solution, the resulting parameters also differ by $<0.1\sigma$ for both functions and the difference between them is still $-2\Delta\ln {\cal L} = 22.6$.
This is because our sample was selected to only include planets with well-determined mass ratios, 
so their uncertainties have little effect.

However, in our sample selection (see above), we also excluded four planets with $\Delta \log q > 0.25$ and three planet candidates with alternative solutions with $\log q > -1.5$ (two other planet candidates with $\log q > -1.5$ solutions do not need to be reconsidered because they also suffer from the 2L1S/1L2S degeneracy, and so are already accounted for in our sensitivity calculation). We reintroduced those planets and planet candidates to our sample and weighted the degenerate solutions by $\Delta\chi^2$ according to the above prescription. We find that the difference between the power-law and double-Gaussian models is almost identical ($-2\Delta\ln {\cal L} = 22.8$). For planet candidates excluded because of the 2L1S/1L2S degeneracy and planets in binary-star systems, those effects were already accounted for in our sensitivity calculation. As discussed above, a correction factor of 11.9\% would account for the ambiguous planets and planet candidates. 

All these tests continue to prefer a double-Gaussian model over a power-law model when fitted to our planet sample.

\newpage
\section*{Supplementary Text}

\subsection{Comparison of methodologies}

Two of the previous works discussed in the main text \cite{Suzuki16,Gould10} used by-eye inspection of the data to identify planets and planet candidates. Those works calculated the planet sensitivity function for their samples by generating and fitting simulated data for planetary events \cite{Rhie00}. The simulated data had the same epochs and uncertainties as the real data, but without noise. Those data were fitted with a PSPL light curve to quantify the magnitude of the planetary signal.

Another work discussed in the main text \cite{Shvartzvald16} used an algorithm to search their data for planetary signals. That algorithm is based on evaluating the contribution of any given subset of the data to the $\chi^2$ of the point lens fit relative to a threshold. That work calculated the planet sensitivity function by simulating data for 2L1S events with the epochs and uncertainties of the real data and adding random noise. They determined the planet sensitivity function by applying the search algorithm to the simulated data.

Similar to that work \cite{Shvartzvald16}, our study uses an algorithm for identifying planetary signals in both the real and simulated data, although the algorithm itself differs. Another difference is that we used the residuals to the best-fitting point lens model to create the simulated data rather than random noise. This allows us to account for correlated noise in our sensitivity calculation.

We observe that the adoption of \textsc{AnomalyFinder} led to a significant increase in the KMTNet planet identifications relative to by-eye searches, particularly at the smallest mass ratios. For the 43 planets in our sample with $\log q > -4$, 29 had previously been identified by eye prior to the \textsc{AnomalyFinder} search (67\%), whereas only nine of the 20 (45\%) $\log q < -4$ planets in our sample had been identified by eye. This contrasts with a previous argument that by-eye detections were relatively unbiased \cite{Gould10}.

These differences in methodologies could contribute to differences in the derived mass-ratio distributions.

\subsection{Comparison to radial velocity studies}

The mass distribution for FGK dwarfs from a radial velocity survey \cite{Mayor11,Fernandes19} showed no evidence of runaway gas accretion for planets with  $10~ M_{\rm Earth} < m_{\rm p} < 100~ M_{\rm Earth}$ \cite{Bennett21}.
For a  $1~ M_{\rm Sun}$ host star, this corresponds to $-4.5 < \log q < -3.5$, mostly below where we identify the change in slope in the cumulative $\log q$ distribution.
The parameter range in that study only partially overlaps with our results, for a single bin of radial velocity periods $\sim 0.7~\mathrm{yr}$ to $\sim 5.3~\mathrm{yr}$.
 
A radial velocity study of M dwarfs found
a low occurrence rate of Jupiter-like planets (defined as similar to Jupiter in mass and received stellar flux) around mid-to-late M dwarfs \cite{Pass23}. For planets in the mass range $0.3 < m_{\rm p} \sin i / M_{\rm Jup} < 10$ (where $M_{\rm Jup}$ is the mass of Jupiter), that study found a planet occurrence rate of $<7.6\% $, which is consistent with our observed frequency of 
$4.6 ^{+1.1}_{-0.8}\%$
over the range $-3 < \log q < -1.5$, which we regard as the equivalent range of mass-ratios.

%%%%%%%%%%%%%
% FIGURES AND TABLES
%%%%%%%%%%%%%

%%%%% FIGURES %%%%%%
\clearpage
\begin{figure}
	\includegraphics[width=\textwidth]{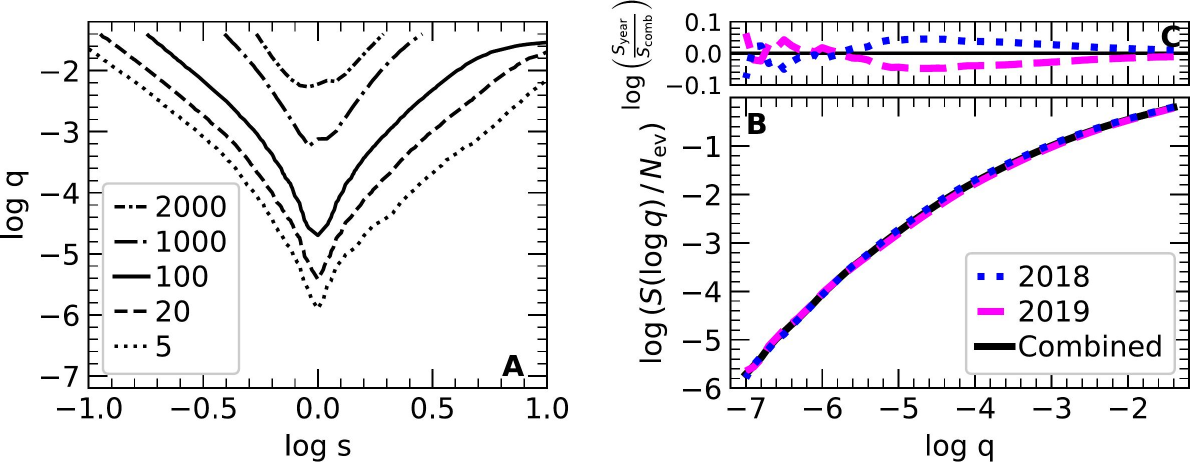}
	\caption{{\bf Sensitivity function of KMTNet. A:} Contours showing the total number of planets that would have been detected in the 2018--2019 KMTNet sample if every star had a planet with a given $\log s$ and $\log q$. Contours are drawn at various values with different line styles as shown in the legend. {\bf B:} Sensitivity as a function of $\log q$, marginalized over $\log s$. The 2018 (blue dotted line), 2019 (magenta dashed line), and combined 2018+2019 (black solid line) curves are nearly identical. {\bf C:} The fractional differences between the combined 2018+2019 curve and the 2018 (blue dotted line) and 2019 (magenta dashed line) sensitivity curves.} 
\end{figure}

% Supplement

\begin{figure*}[htb]
    \centering
    \includegraphics[width=0.49\linewidth]{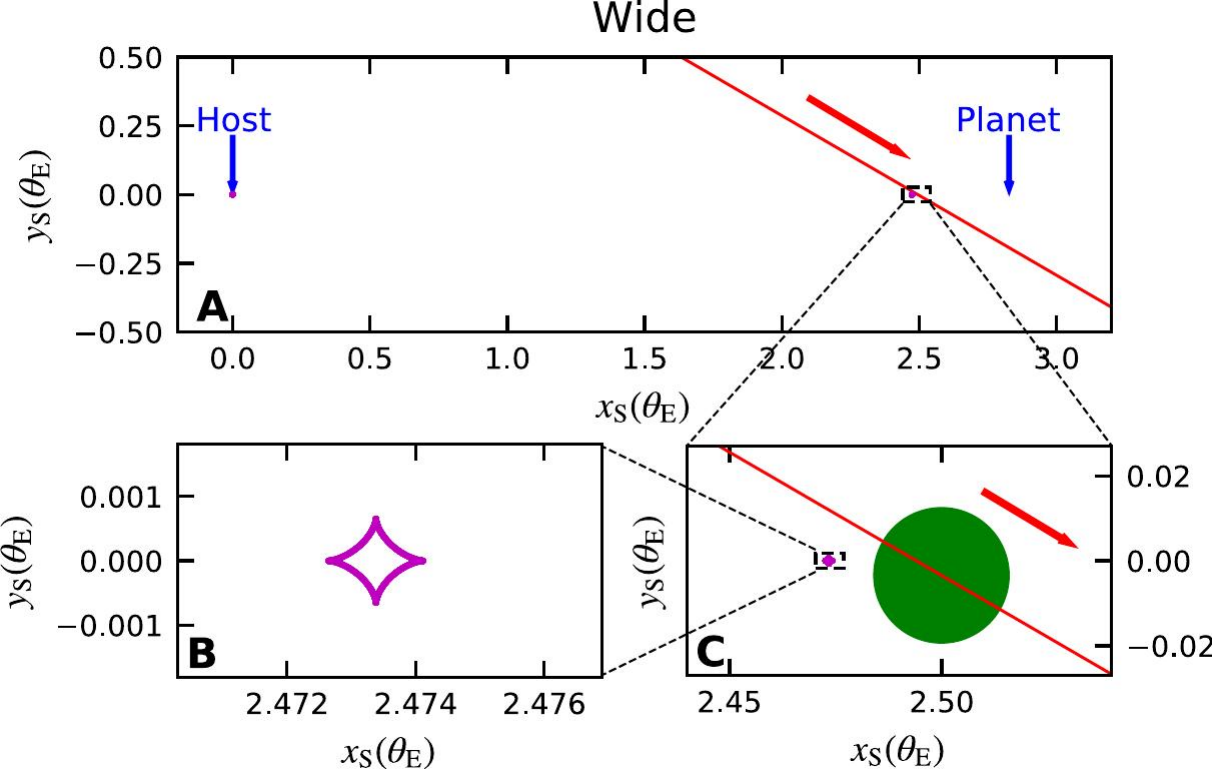}
    \includegraphics[width=0.49\linewidth]{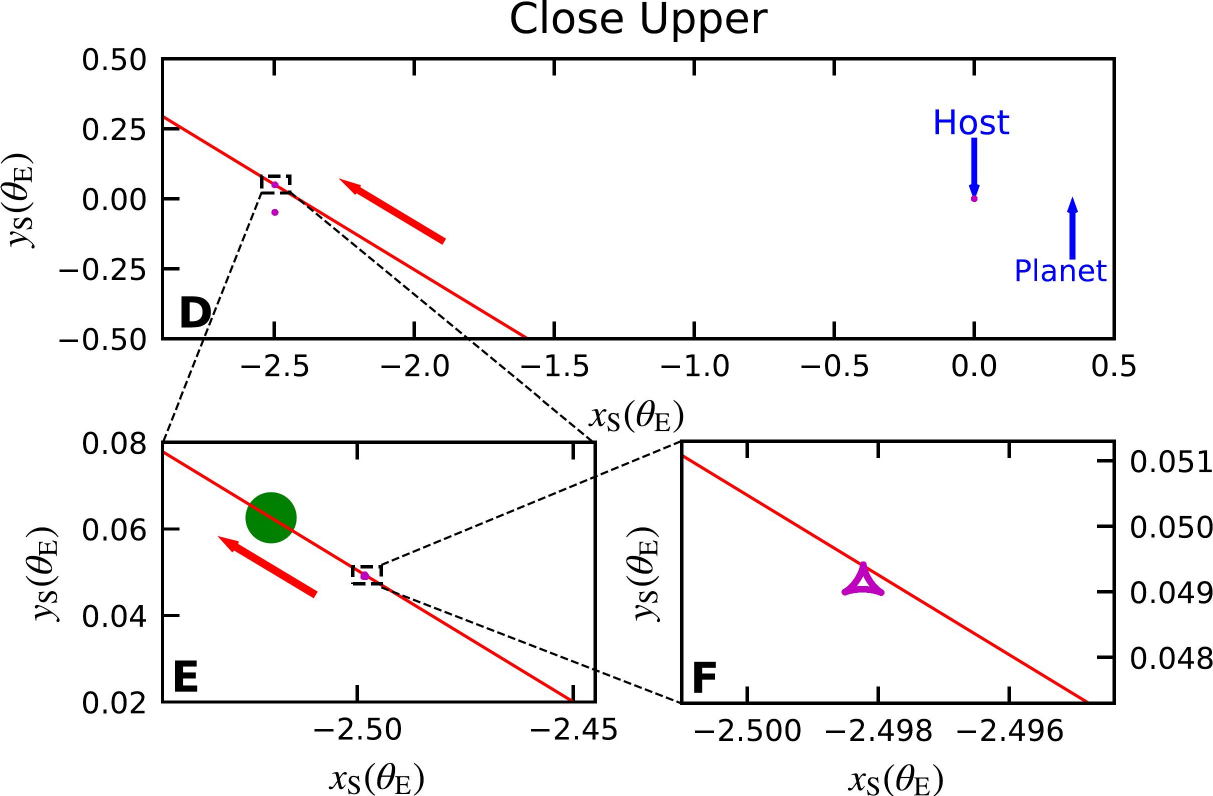}
    \includegraphics[width=0.49\linewidth]{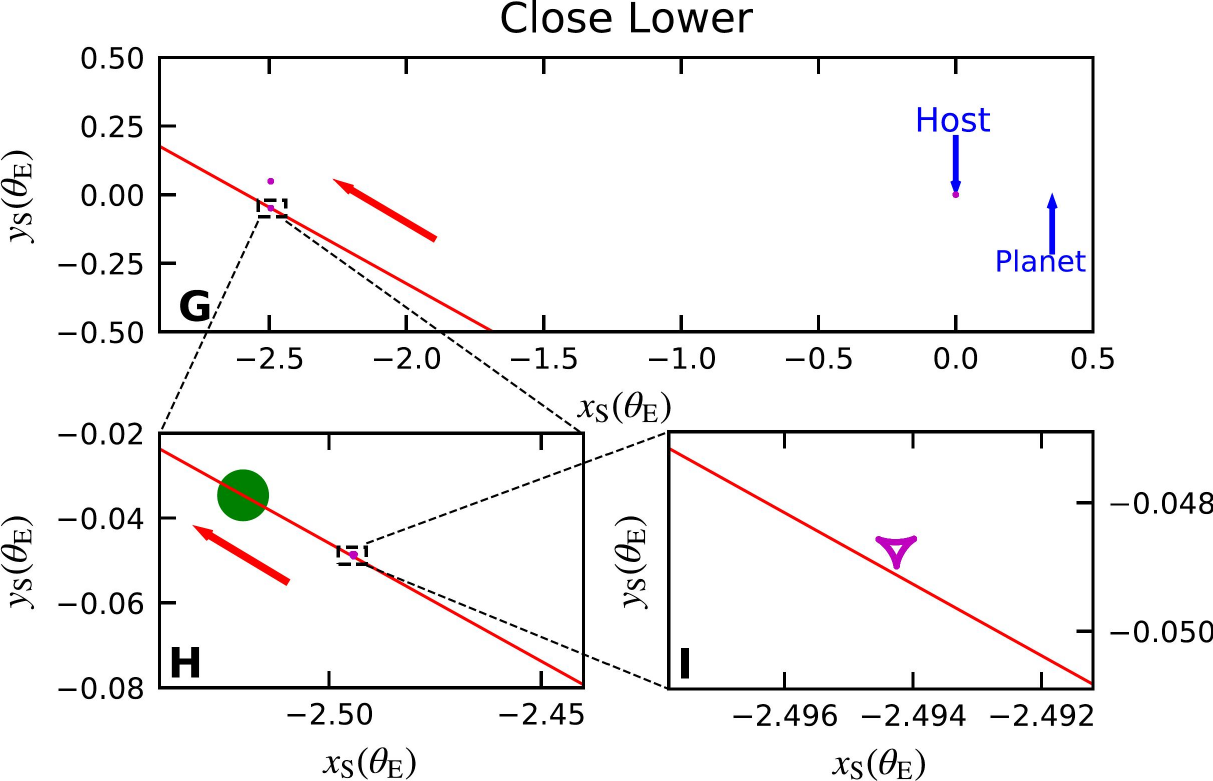}
    \caption{{\bf Geometries of three potential solutions for OGLE-2016-BLG-0007.} These are shown in the plane of the source, scaled to the Einstein radius ($\theta_{\rm E}$), with the $x_{\rm S}$ axis aligned with the axis connecting the host star and planet and the $y_{\rm S}$ axis in the perpendicular direction. The locations of the host star and planet are indicated by the blue arrows and the resulting caustic structure is shown in magenta. The trajectory of the source relative to the lens is shown as the red line with an arrow indicating its direction. {\bf A--C:} The Wide solution,  which we conclude is the most probable for this event; {\bf D--F:} The Close Upper and {\bf G--I:} Close Lower solutions are excluded by the light-curve modeling. The top panel in each set of three panels shows the full caustic structure, the host star, and the planet. The bottom panels show a zoom on the source to demonstrate its relative size and a zoom on the planetary caustic to show its detailed structure. The green circle shows the source star to scale at its location just after the planetary anomaly (HJD$^{\prime} = 7658.8$ for panel {\bf C} and HJD$^{\prime} = 7658.4$ for panels {\bf E} and {\bf H}).}
\end{figure*}

\begin{figure}[htb] 
    \centering
    \includegraphics[height=0.7\textheight]{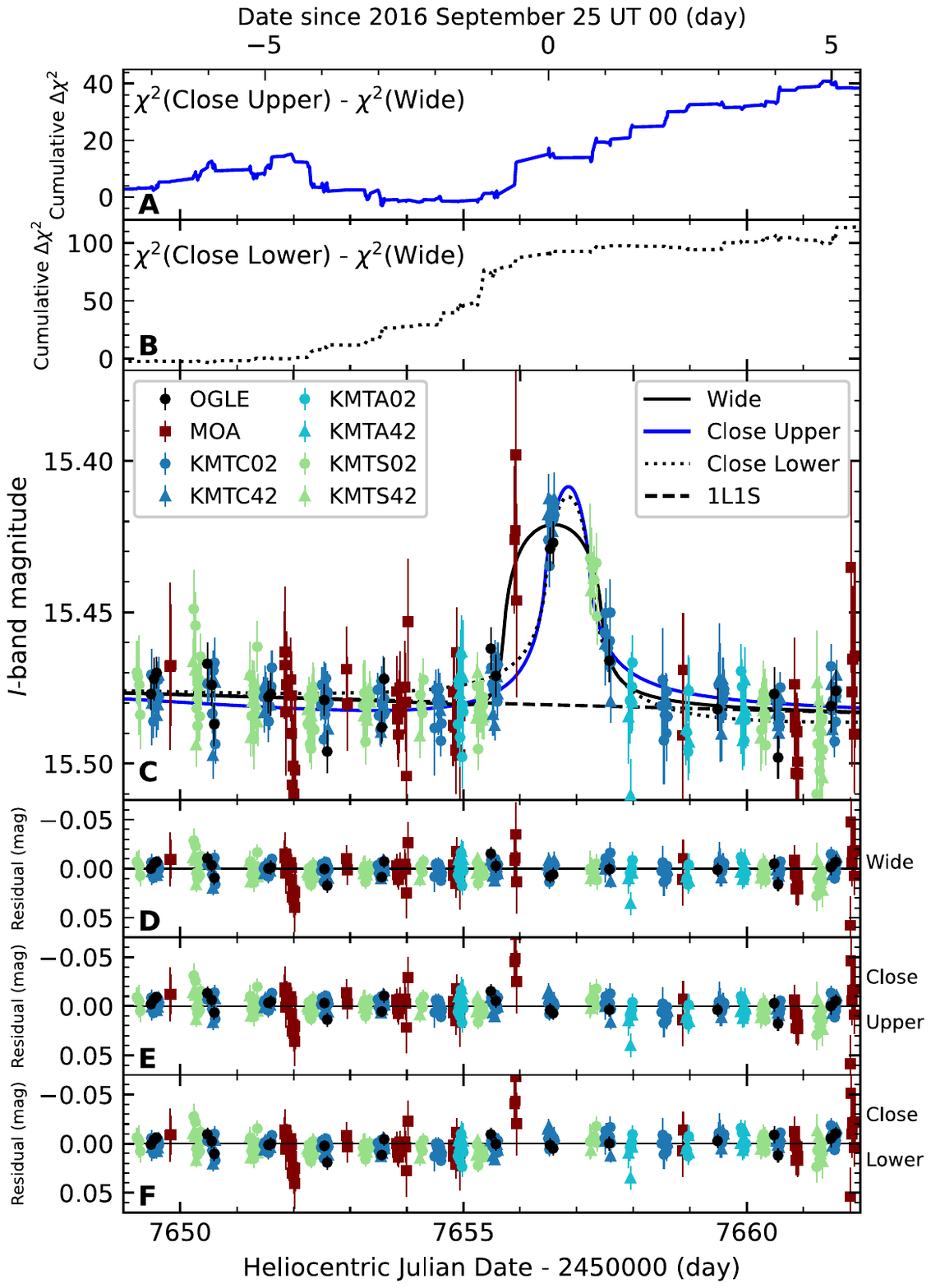}
    \caption{{\bf Comparison of models fitted to the light curve of OGLE-2016-BLG-0007.} {\bf A:} the cumulative $\chi^2$ between the best-fitting solution and the Close Upper solution.
  {\bf B:} Same as {\bf A} but for the Close Lower solution.
  {\bf C:} The data (see legend) and fitted models over the same time range. The best-fitting solution is the Wide solution (black line), while the Close Upper and Close Lower solutions are significantly disfavored by the data. The 1L1S model corresponding to the Wide model but without the planet is shown as the dashed gray line to demonstrate the effect of the planet on the light curve. {\bf D--F:} Residuals between the data and the models. 
  The preference for the Wide solution comes from datapoints from multiple observatories and fields taken during the anomaly.
  The data are shown with $1\sigma$ error bars.}
\end{figure}

\begin{figure}[htb]
    \centering
    \includegraphics[width=0.7\textheight]{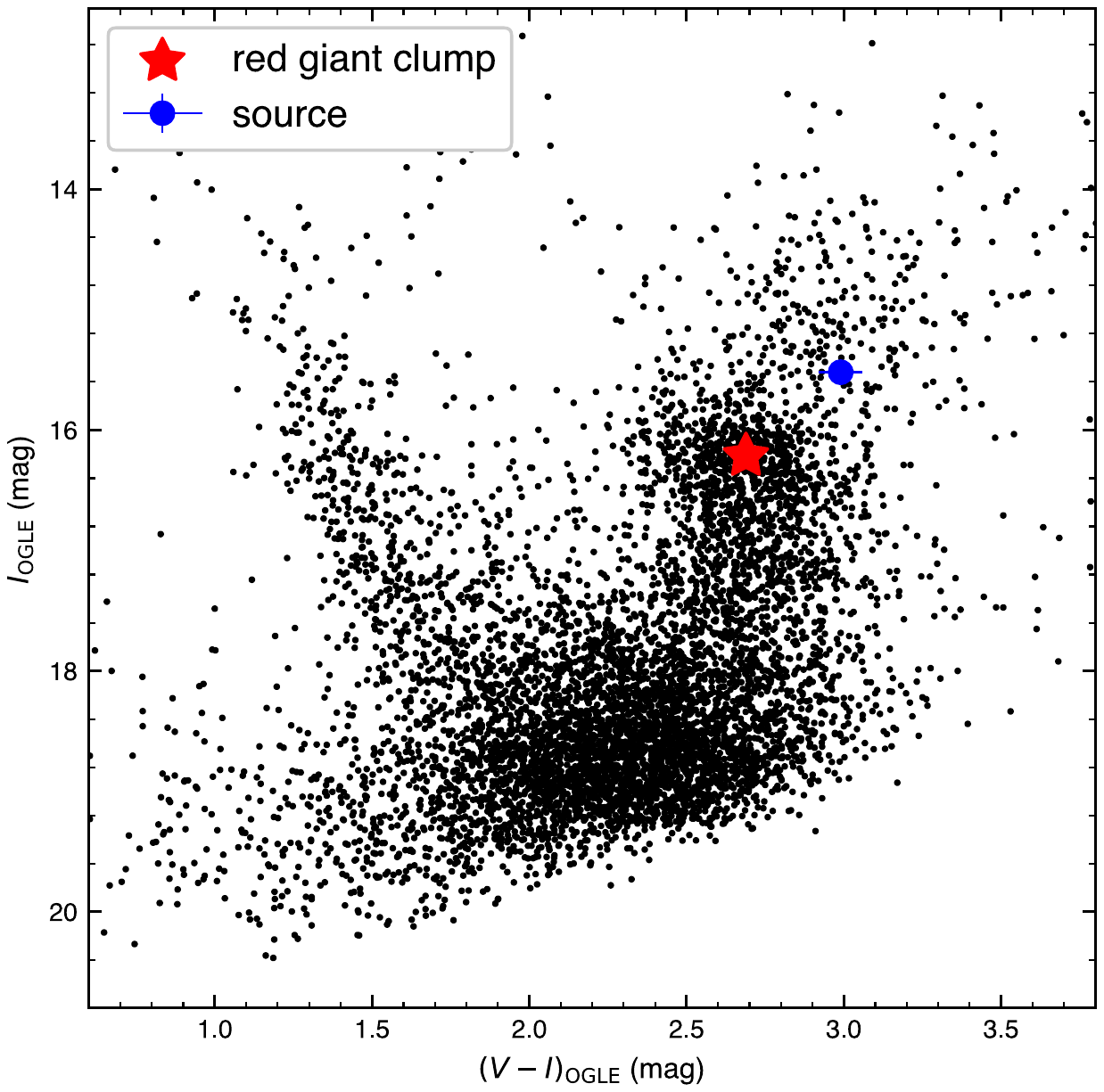}
     \caption{{\bf Color-Magnitude Diagram of stars around OGLE-2016-BLG-0007.} Black points are data for stars in the OGLE-III catalog \cite{OGLEIII} within $120^{\prime\prime}$ of OGLE-2016-BLG-0007. The red star is the centroid of the red clump, and the blue circle shows the location of the source star with $1\sigma$ error bars.}
\end{figure}

\begin{figure}
	\centering
	\includegraphics[height=0.7\textheight]{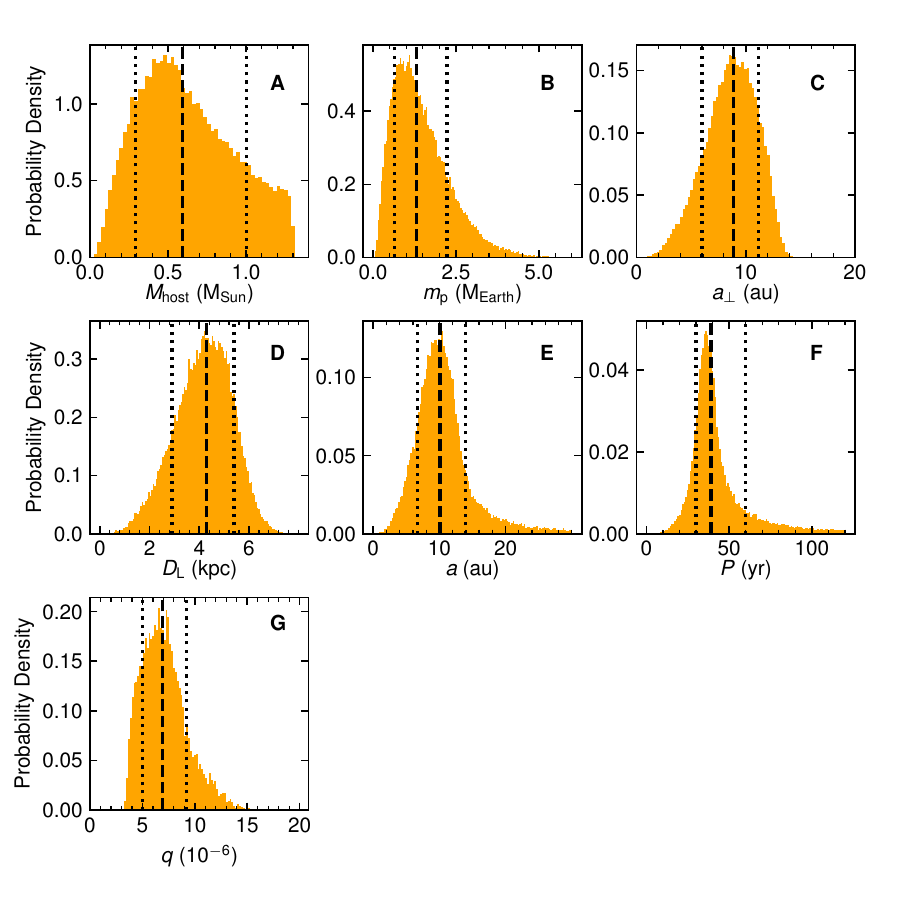}
	\caption{{\bf Posterior probability distributions for the lens system properties} derived from the MCMC analysis (orange). The black lines indicate the median (dashed line) and 68\% confidence interval (dotted lines). The primary physical properties of the the lens are the host mass $M_{\rm host}$ ({\bf A}), planet mass $m_{\rm p}$ ({\bf B}), projected host-planet separation $a_\bot$ ({\bf C}) and the distance to the lens system $D_{\rm L}$ ({\bf D}). The posteriors for the semi-major axis $a$ ({\bf E}) and period $P$ ({\bf F}) were derived assuming circular orbits. Also shown is the posterior probability for the mass-ratio $q$ ({\bf G}).}
\end{figure}

\begin{figure}
	\centering
	\includegraphics[height=0.6\textheight]{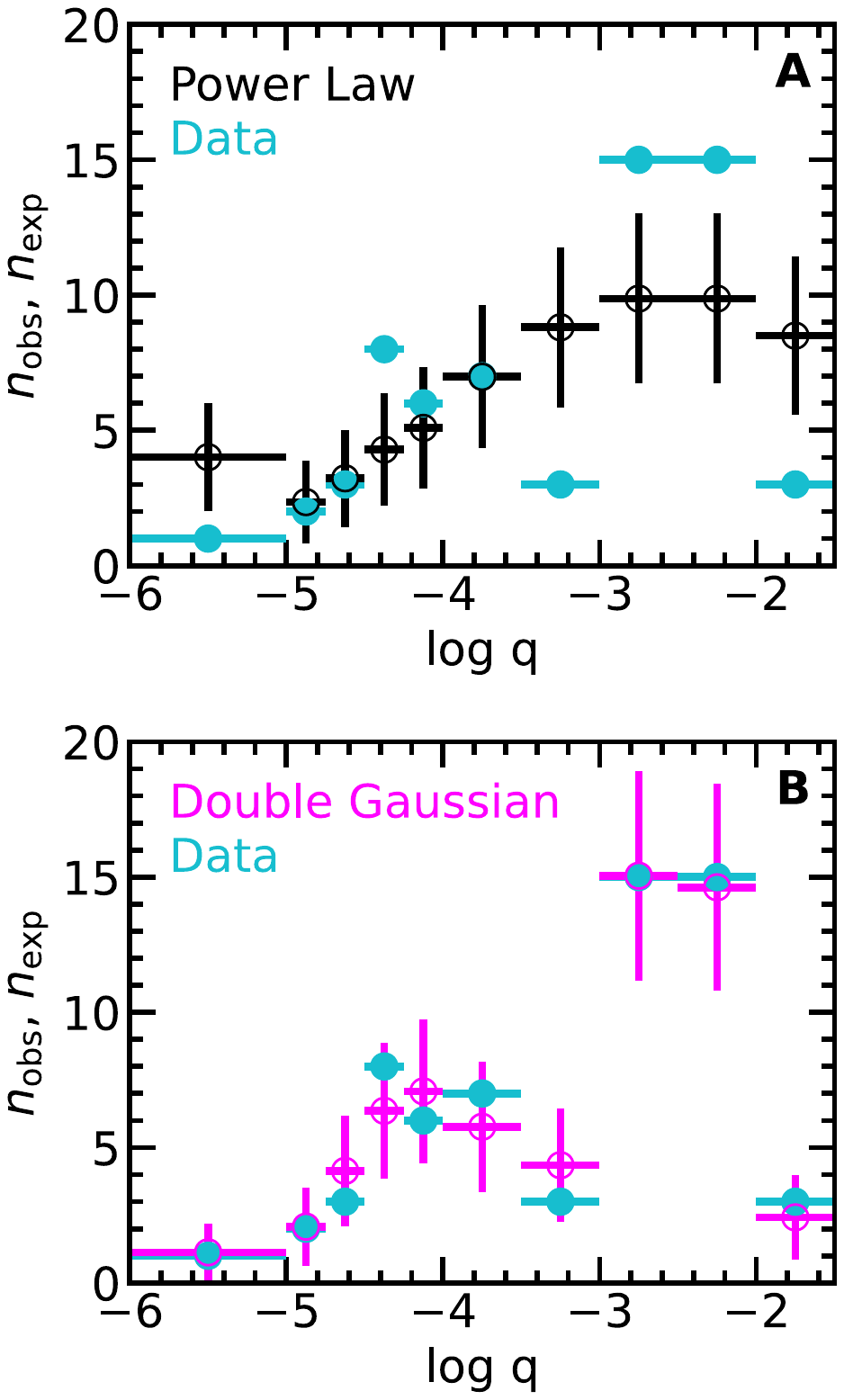}
	\caption{{\bf Comparison of observed and predicted numbers of planet detections.} The cyan points show the number of observed planet detections, $n_{\rm obs}$, for bins of $\log~q$ with  various widths indicated by the horizontal errorbars.  The values predicted by the best-fitting models corrected by the sensitivity function, $n_{\rm exp}$, are shown in black for the power-law model ({\bf A}) and in magenta for the double Gaussian model ({\bf B}). The uncertainties in the model values are Poisson and are approximated to be $\sqrt{n_{\rm exp}}$ (vertical errorbars); see Fig. S7 for a more accurate representation of the Poisson distribution function for a subset of these bins.}
\end{figure}

\begin{figure}
	\centering
	\includegraphics[height=0.8\textheight]{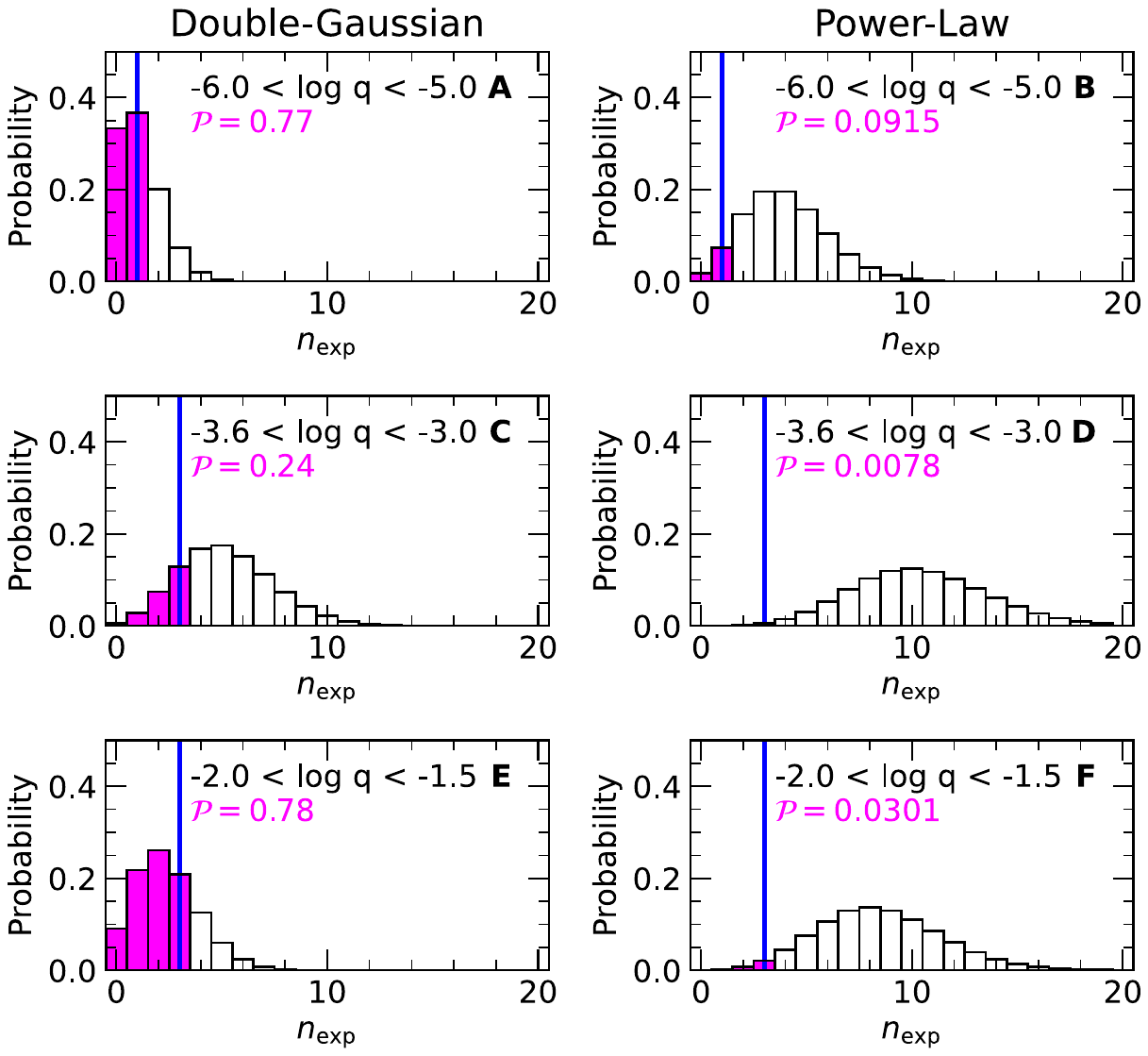}
	\caption{{\bf Poisson probabilities for the expected number of planets}, $n_{\rm exp}$, in each mass-ratio bin for which the observed number is significantly smaller than the prediction of the power-law model. {\bf A}, {\bf C}, and {\bf E} show the Poisson distributions (black bars) for the expected number of planets in each $\log q$ bin (labeled  in each panel) for the best-fitting double-Gaussian model compared to the observed number of planets (blue line). The magenta-shaded regions indicate cases with the same or fewer planets detected. The total probability of the magenta regions, ${\cal P}$, is labeled in magenta in each panel. 
	{\bf B}, {\bf D}, and {\bf F} are the same except for the best-fitting power-law model.}
\end{figure}

\begin{figure}
	\centering
	\includegraphics{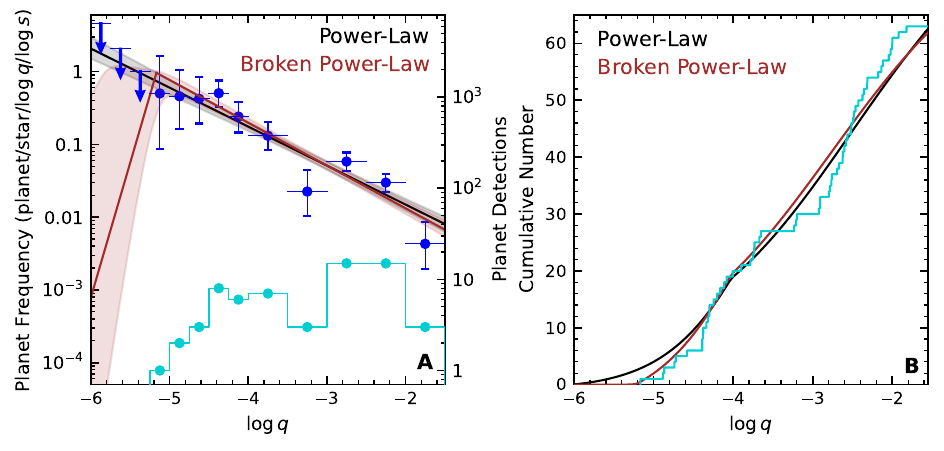}
	\caption{{\bf Comparison of the broken power law model to the inferred planet mass ratio distribution.} Same as Fig. 3, except showing the best-fitting broken power-law model (brown) instead of the double Gaussian model. The broken power-law model is not significantly different from the single power-law model.}
\end{figure}

\begin{figure}
	\centering
	\includegraphics[width=\linewidth]{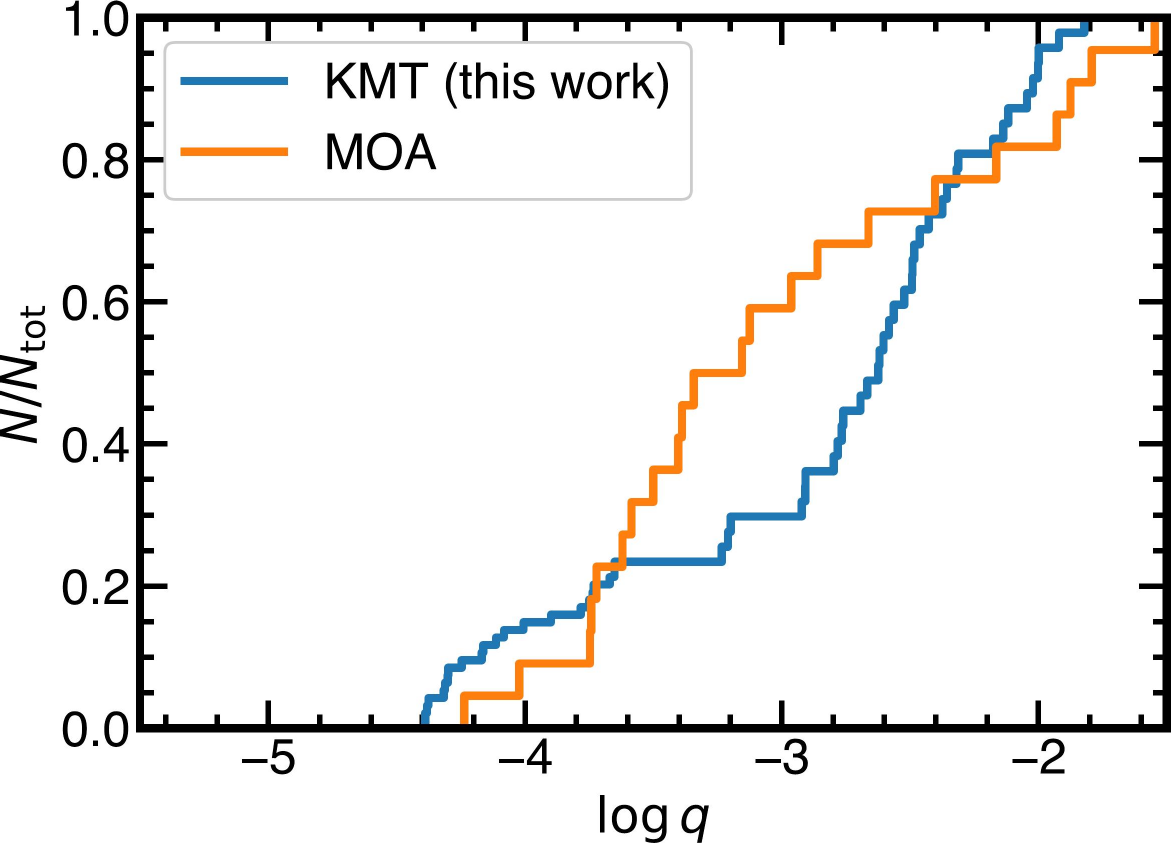}
	\caption{{\bf Fractional cumulative distributions of planets with $\log q > -4.5$ as a function of $\log q$.} Our planet sample (blue line) is compared to the previously published MOA planet sample \cite{Suzuki16} (orange). KMT planets with $\log q < -4$ are drawn from four seasons, rather than two (as for $\log q > -4$) and have been weighted accordingly.} 
\end{figure}

%%%% TABLES %%%%%
%\FloatBarrier
\clearpage
%\renewcommand\arraystretch{1.25}
%\footnotesize
\setlength\tabcolsep{2pt}
    \begin{center}
    \resizebox{\columnwidth}{!}{\begin{tabular}{c | c c c || c | c}
    \multicolumn{5}{p{\linewidth}}{{\bf Table S1. Fitted parameters for four potential models of the OGLE-2016-BLG-0007 light curve.} The Wide solution is preferred by the data. The uncertainties are the 68\% confidence intervals centered on the median from the MCMC fits.}\\
    \hline
    \hline
    Parameter & \multicolumn{3}{c||}{2L1S} & Parameter & 1L2S \\
    & Wide & Close Upper & Close Lower & \\
    \hline
    $\chi^2$/degrees of freedom & $10777.0/10777$ & $10815.5/10777$ & $10892.2/10777$ & & $10922.6/10776$ \\
    \hline
    $t_{0}$ (${\rm HJD}^{\prime}$) & $7498.44 \pm 0.10$ & $7498.44 \pm 0.10$ & $7498.46 \pm 0.11$ & $t_{0,1}$ (${\rm HJD}^{\prime}$) & $7498.43 \pm 0.12$ \\
    ... & ... & ... & ... & $t_{0,2}$ (${\rm HJD}^{\prime}$) &$7656.824 \pm 0.022$ \\
    $u_{0}$ & $1.2532 \pm 0.0033$ & $1.2469 \pm 0.0037$ & $1.2568 \pm 0.0033$ & $u_{0,1}$ & $1.2549 \pm 0.0033$ \\
    ... & ... & ... & ... & $u_{0,2}$ ($10^{-4}$) & $0.0\pm 4.8$ \\
    $t_{\rm E}$ (days) & $73.92 \pm 0.31$ & $73.37 \pm 0.31$ & $73.55 \pm 0.31$ & $t_{\rm E}$ (days) & $74.96 \pm 0.37$ \\
    $\rho$ ($10^{-2}$) & $1.51 \pm 0.17$ & $0.554 \pm 0.038$ & $0.552 \pm 0.028$ & $\rho_2$ ($10^{-2}$) & $0.479 \pm 0.034$ \\
     ... & ... & ... & ... & $q_{f,I} (10^{-4})$ & $1.62 \pm 0.10$ \\
     ... & ... & ... & ... & $q_{f, \rm MOA} (10^{-4})$ &$0.38 \pm 0.29$ \\
    $\alpha$ (rad) & $5.7565 \pm 0.0016$ & $2.5943 \pm 0.0014$ & $2.6336 \pm 0.0015$ & ... & ... \\
    $s$ & $2.8287 \pm 0.0088$ & $0.3510 \pm 0.0010$ & $0.3516 \pm 0.0010$ & ... & ... \\
    $\log q$ & $-5.17 \pm 0.13$ & $-4.063 \pm 0.032$ & $-4.087 \pm 0.027$ & ... & ... \\
    $I_{\rm S}$ & $15.5213 \pm 0.0063$ & $15.5208 \pm 0.0063$ & $15.521 \pm 0.0063$ &$I_{\rm S}$ & $15.523 \pm 0.0063$ \\
    \hline
    \hline
    \end{tabular}}
    \end{center}

% NEW
    \begin{center}
    \begin{tabular}{p{0.645\textwidth} p{0.225\textwidth} p{0.13\textwidth}}
    \multicolumn{3}{p{\textwidth}}{{\bf Table S2. Colors, magnitudes, and other physical properties measured, fitted, or derived for OGLE-2016-BLG-0007.} Values are given with $1\sigma$ uncertainties.}\\
    \hline
    \hline    
    \multicolumn{3}{l}{Colors} \\    
    \hline
    Parameter &  Symbol (units) &  Value\\
    \hline
    Color of red clump centroid from CMD & $(V - I)_{\rm cl}$ (mag)& $2.6863 \pm 0.0052$  \\
    Intrinsic color of red clump centroid \cite{Bensby2013}  & $(V - I)_{\rm cl, 0}$  (mag) & $1.060 \pm 0.030$  \\
    Source color from model fitting & $(V - I)_{\rm S}$  (mag) &  $3.059 \pm 0.072$ \\    
    Derived intrinsic source color & $(V - I)_{\rm S,0}$ (mag)  &$1.433 \pm 0.078$ \\
    \hline
    \hline
    \multicolumn{3}{l}{Magnitudes} \\
    \hline
    Parameter &  Symbol (units) &  Value\\
    \hline
    Apparent magnitude of red clump centroid from CMD &$I_{\rm cl}$ (mag) &  $16.220 \pm 0.011$ \\
    Expected intrinsic apparent magnitude of red clump centroid \cite{Nataf2013} &$I_{\rm cl, 0}$ (mag)  & $14.430 \pm 0.040$ \\
    Apparent magnitude of source from model fitting &$I_{\rm S}$ (mag)  & $15.5213 \pm 0.0063$  \\
    Derived intrinsic apparent magnitude of source & $I_{\rm S,0}$ (mag)  & $13.731 \pm 0.042$ \\
    \hline
    \hline
    \multicolumn{3}{l}{Physical Properties}\\
    \hline
    Parameter &  Symbol (units) &  Value\\
    \hline
    Angular radius of the source star from $(V-I, I)_{\rm S, 0}$ and surface brightness relations \cite{Adams2018} & $\theta_*$ (microarcseconds)&$11.0 \pm 1.0$\\
    Derived angular Einstein radius & $\theta_{\rm E}$ (mas) & $0.73 \pm 0.10$\\
    Derived lens-source relative proper motion  & $\mu_{\rm rel}$ (${\rm mas~yr^{-1}}$) & $3.61 \pm 0.51$\\
    \hline
    \hline
    \end{tabular}
    \end{center}

%\newpage
\renewcommand\arraystretch{1.0}
\begin{center}  
    \begin{tabular}{l c c c l}
    \multicolumn{5}{p{\textwidth}}{{\bf Table S3. Microlensing planets selected for our statistical sample.} Entries are listed by the mean $\log q$ of the solutions. The Event Name is the primary designation for each event based on which survey identified the event first. The KMTNet Name is the designation assigned by the KMTNet Collaboration. Both the event names and KMTNet names are presented in abbreviated format: SBYYNNNN, where S is the survey that identified the event (O = OGLE, and K= KMT), B indicates that the event was identified in the Galactic Bulge fields observed by that survey, YY is the 2-digit abbreviation for the year (e.g. 16 = 2016), and NNNN is the 4-digit number assigned to the event by the survey in the order that the events were identified within a year. For each planet, we give the $\log{q}$ and $s$ of each degenerate solution and the reference(s) those solutions were taken from.}\\
    \hline
    \hline
    Event Name & KMTNet Name & $\log{q}$ & $s$ & Reference \\
    \hline
    OB160007 & KB161991 & $-5.168 \pm 0.131$ & $2.8287 \pm 0.0088$ & This work \\
    \hline
    OB191053 & KB191504 & $-4.885 \pm 0.035$ & $1.406 \pm 0.011$ & \cite{Zang21AF1} \\
    \hline
    OB190960 & KB191591 & $-4.830 \pm 0.041$ & $1.029 \pm 0.001$ & \cite{OB190960} \\
    & & $-4.896 \pm 0.024$ & $0.997 \pm 0.001$ & ... \\
    & & $-4.896 \pm 0.024$ & $0.996 \pm 0.001$ & ... \\
    & & $-4.845 \pm 0.043$ & $1.028 \pm 0.001$ & ... \\
    \hline
    KB180029 & KB180029 & $-4.738 \pm 0.052$ & $1.000 \pm 0.002$ & \cite{KB180029} \\
    & & $-4.755 \pm 0.055$ & $1.000 \pm 0.002$ & ... \\   
    \hline
    KB191806 & KB191806 & $-4.714 \pm 0.116$ & $1.035 \pm 0.009$ & \cite{Zang23AF7} \\
    & & $-4.717 \pm 0.117$ & $1.034 \pm 0.009$ & ... \\  
    & & $-4.724 \pm 0.117$ & $0.938 \pm 0.007$ & ... \\  
    & & $-4.734 \pm 0.109$ & $0.938 \pm 0.007$ & ... \\  
    \hline
    KB171194 & KB171194 & $-4.582 \pm 0.058$ & $0.806 \pm 0.010$ & \cite{Zang23AF7} \\
    \hline
    KB190842 & KB190842 & $-4.389 \pm 0.031$ & $0.983 \pm 0.013$ & \cite{KB190842} \\
    \hline
    KB190253 & KB190253 & $-4.387 \pm 0.076$ & $1.009 \pm 0.009$ & \cite{Hwang22AF2} \\
    & & $-4.390 \pm 0.080$ & $0.929 \pm 0.007$ & ... \\
    \hline
    OB180977 & KB180728 & $-4.382 \pm 0.045$ & $0.897 \pm 0.007$ & \cite{Hwang22AF2} \\
    \hline
    OB171806 & KB171021 & $-4.352 \pm 0.171$ & $0.857 \pm 0.008$ & \cite{Zang23AF7} \\
    & & $-4.392 \pm 0.180$ & $0.861 \pm 0.007$ & ... \\
    & & $-4.441 \pm 0.168$ & $1.181 \pm 0.011$ & ... \\
    & & $-4.317 \pm 0.126$ & $1.190 \pm 0.012$ & ... \\
    \hline
    OB161195 & KB160372 & $-4.339 \pm 0.037$ & $0.991 \pm 0.004$ & \cite{OB161195,OB161195_MOA,OB161195_true} \\
    & & $-4.337 \pm 0.037$ & $1.076 \pm 0.005$ & ... \\
    \hline
    KB171003 & KB171003 & $-4.373 \pm 0.144$ & $0.910 \pm 0.005$ & \cite{Zang23AF7} \\
    & & $-4.260 \pm 0.152$ & $0.889 \pm 0.004$ & ... \\
    \hline
    KB191367 & KB191367 & $-4.303 \pm 0.118$ & $0.939 \pm 0.007$ & \cite{Zang23AF7} \\
    & & $-4.298 \pm 0.103$ & $0.976 \pm 0.007$ & ... \\
    \hline
    KB170428 & KB170428 & $-4.295 \pm 0.072$ & $0.882 \pm 0.004$ & \cite{Zang23AF7} \\
    & & $-4.302 \pm 0.075$ & $0.915 \pm 0.005$ & ... \\
     \hline
    \hline
   \end{tabular}

% break table here    
    \begin{tabular}{l c c c l}
    \multicolumn{5}{c}{Table S3. (cont.)}\\
    \hline
    \hline
        Event Name & KMTNet Name & $\log{q}$ & $s$ & Reference \\
   \hline
    OB171434 & KB170016 & $-4.242 \pm 0.011$ & $0.979 \pm 0.001$ & \cite{Udalski18} \\
    & & $-4.251 \pm 0.012$ & $0.979 \pm 0.001$ & ... \\
    \hline
    KB181025 & KB181025 & $-4.202 \pm 0.137$ & $0.943 \pm 0.021$ & \cite{KB181025}\\
    \hline
    OB181185 & KB181024 & $-4.163 \pm 0.014$ & $0.963 \pm 0.001$ & \cite{OB181185} \\
    \hline
    OB180506 & KB180835 & $-4.117 \pm 0.133$ & $1.059 \pm 0.021$ & \cite{Hwang22AF2} \\
    & & $-4.109 \pm 0.126$ & $0.861 \pm 0.018$ & ... \\
    \hline
    OB171691 & KB170752 & $-4.013 \pm 0.152$ & $1.003 \pm 0.014$ & \cite{OB171691} \\
    & & $-4.150 \pm 0.141$ & $1.058 \pm 0.011$ & ... \\
    \hline
    OB180532 & KB181161 & $-3.994 \pm 0.024$ & $1.012 \pm 0.001$ & \cite{OB180532} \\
    & & $-4.017 \pm 0.025$ & $1.013 \pm 0.001$ & ... \\
    \hline
    OB180516 & KB180808 & $-3.880 \pm 0.046$ & $1.006 \pm 0.005$ & \cite{Hwang22AF2} \\
    & & $-3.918 \pm 0.045$ & $0.867 \pm 0.004$ & ... \\
    \hline
    OB180298 & KB181354 & $-3.705 \pm 0.099$ & $0.957 \pm 0.018$ & \cite{Jung22AF6} \\
    & & $-3.861 \pm 0.081$ & $1.079 \pm 0.016$ & ... \\
    \hline
    KB190953 & KB190953 & $-3.750 \pm 0.083$ & $0.736 \pm 0.010$ & \cite{Hwang22AF2} \\
    \hline
    OB191492 & KB193004 & $-3.719 \pm 0.132$ & $0.898 \pm 0.015$ & \cite{Hwang22AF2} \\
    & & $-3.754 \pm 0.133$ & $1.044 \pm 0.018$ & ... \\
    \hline
    OB180596 & KB180945 & $-3.738 \pm 0.031$ & $0.512 \pm 0.017$ & \cite{OB180596} \\
    & & $-3.726 \pm 0.031$ & $0.499 \pm 0.018$ & ... \\
    \hline
    OB180383 & KB180900 & $-3.670 \pm 0.069$ & $2.453 \pm 0.026$ & \cite{OB180383} \\
    \hline
    KB191216 & KB191216 & $-3.644 \pm 0.116$ & $1.083 \pm 0.013$ & \cite{Jung23AF8} \\
    & & $-3.658 \pm 0.120$ & $1.037 \pm 0.022$ & ... \\
    & & $-3.661 \pm 0.121$ & $1.086 \pm 0.016$ & ... \\
    & & $-3.639 \pm 0.114$ & $1.047 \pm 0.023$ & ... \\
    \hline
    OB181269 & KB182418 & $-3.225 \pm 0.017$ & $1.124 \pm 0.033$ & \cite{OB181269} \\
    & & $-3.240 \pm 0.020$ & $1.123 \pm 0.032$ & ... \\
    \hline
    KB192974 & KB192974 & $-3.209 \pm 0.076$ & $0.854 \pm 0.011$ & \cite{Zang22AF4} \\
    \hline
    KB191042 & KB191042 & $-3.201 \pm 0.041$ & $1.017 \pm 0.004$ & \cite{Zang22AF4} \\
    & & $-3.194 \pm 0.043$ & $1.125 \pm 0.004$ & ... \\
    \hline
    OB180932 & KB182087 & $-2.922 \pm 0.026$ & $0.536 \pm 0.001$ & \cite{Gould22AF5} \\
    \hline
    OB181212 & KB182299 & $-2.908 \pm 0.015$ & $1.451 \pm 0.016$ & \cite{Gould22AF5} \\
     & & $-2.909 \pm 0.015$ & $0.680 \pm 0.007$ & ... \\
    \hline
    OB180567 & KB180890 & $-2.907 \pm 0.024$ & $1.806 \pm 0.019$ & \cite{OB180567} \\
    \hline
    KB181996 & KB181996 & $-2.821 \pm 0.112$ & $1.455 \pm 0.091$ & \cite{Han2021three} \\
    & & $-2.772 \pm 0.098$ & $0.672 \pm 0.041$ & ... \\
    \hline
    KB182602 & KB182602 & $-2.782 \pm 0.071$ & $1.182 \pm 0.065$ & \cite{Jung22AF6} \\
    \hline
    OB181428 & KB180423 & $-2.766 \pm 0.014$ & $1.423 \pm 0.002$ & \cite{OB181428} \\
    \hline
    OB181119 & KB181870 & $-2.74 \pm 0.11$ & $1.426 \pm 0.047$ & \cite{Jung22AF6} \\
    & & $-2.78 \pm 0.11$ & $1.081 \pm 0.038$ & ... \\
    \hline
     \hline
   \end{tabular}

% break table here    
    \begin{tabular}{l c c c l}
    \multicolumn{5}{c}{Table S3. (cont.)}\\
    \hline
    \hline
        Event Name & KMTNet Name & $\log{q}$ & $s$ & Reference \\
    \hline
    KB180748 & KB180748 & $-2.693 \pm 0.032$ & $0.939 \pm 0.003$ & \cite{KB180748} \\
    \hline
    OB190954$^{\dagger}$ & KB193289 & $-2.667 \pm 0.056$ & $0.710 \pm 0.005$ & \cite{Zang22AF4} \\
    \hline
    OB180962 & KB182071 & $-2.624 \pm 0.014$ & $1.246 \pm 0.027$ & \cite{OB180567} \\
    \hline
    KB180087 & KB180087 & $-2.668 \pm 0.092$ & $0.898 \pm 0.024$ & \cite{Jung22AF6} \\
    & & $-2.569 \pm 0.083$ & $0.638 \pm 0.014$ & ... \\
    \hline
    KB190298 & KB190298 & $-2.603 \pm 0.059$ & $1.892 \pm 0.030$ & \cite{Jung23AF8} \\
    \hline
    OB180799 & KB181741 & $-2.582 \pm 0.023$ & $1.116 \pm 0.008$ & \cite{OB180799} \\
    \hline
    KB180030 & KB180030 & $-2.563 \pm 0.048$ & $1.580 \pm 0.013$ & \cite{Jung22AF6} \\
    \hline
    KB181976 & KB181976 & $-2.504 \pm 0.132$ & $1.227 \pm 0.064$ & \cite{Han2021three} \\
    & & $-2.539 \pm 0.119$ & $0.708 \pm 0.030$ & ... \\
    \hline
    OB191180 & KB191912 & $-2.480 \pm 0.072$ & $1.867 \pm 0.010$ & \cite{Chung23_ob1180}\\
    & & $-2.465 \pm 0.063$ & $1.874 \pm 0.011$ & ... \\
    \hline
    OB181367 & KB180914 & $-2.48 \pm 0.12$ & $0.566 \pm 0.044$ & \cite{Gould22AF5} \\
    & & $-2.50 \pm 0.13$ & $1.71 \pm 0.14$ & ... \\
    \hline
    KB192783 & KB192783 & $-2.483 \pm 0.101$ & $0.814 \pm 0.007$ & \cite{Jung23AF8} \\
    \hline
    KB181292 & KB181292 & $-2.450 \pm 0.033$ & $1.366 \pm 0.010$ & \cite{KB181292} \\
    & & $-2.474 \pm 0.037$ & $1.355 \pm 0.010$ & ... \\
    \hline
    OB190468$^{*}$ & KB192696 & $-2.451 \pm 0.022$ & $0.853 \pm 0.003$ & \cite{OB190468} \\
    & & $-2.480 \pm 0.026$ & $0.854 \pm 0.004$ & ... \\
    \hline
    KB181990 & KB181990 & $-2.452 \pm 0.044$ & $0.963 \pm 0.007$ & \cite{KB181990} \\
    & & $-2.438 \pm 0.044$ & $0.960 \pm 0.007$ & ... \\
    & & $-2.378 \pm 0.050$ & $1.095 \pm 0.010$ & ... \\
    & & $-2.380 \pm 0.044$ & $1.099 \pm 0.009$ & ... \\
    & & $-2.423 \pm 0.035$ & $0.963 \pm 0.005$ & ... \\
    & & $-2.426 \pm 0.029$ & $0.963 \pm 0.003$ & ... \\
    & & $-2.245 \pm 0.026$ & $1.118 \pm 0.009$ & ... \\
    & & $-2.253 \pm 0.034$ & $1.115 \pm 0.012$ & ... \\
    \hline
    OB180740 & KB181822 & $-2.343 \pm 0.049$ & $1.26 \pm 0.01$ & \cite{OB180740} \\
    & & $-2.369 \pm 0.042$ & $0.86 \pm 0.01$ & ... \\
    \hline
    OB190679 & KB192688 & $-2.319 \pm 0.013$ & $2.216 \pm 0.023$ & \cite{Jung23AF8} \\
    \hline
    KB191552 & KB191552 & $-2.333 \pm 0.049$ & $0.780 \pm 0.007$ & \cite{Zang22AF4} \\
    & & $-2.290 \pm 0.050$ & $0.776 \pm 0.006$ & ... \\
    \hline
    KB180247 & KB180247 & $-2.149 \pm 0.034$ & $1.118 \pm 0.005$ & \cite{Jung22AF6} \\
    & & $-2.203 \pm 0.032$ & $0.972 \pm 0.005$ & ... \\
    \hline
    OB190362 & KB190075 & $-2.126 \pm 0.089$ & $0.898 \pm 0.021$ & \cite{OB190362} \\
    & & $-2.148 \pm 0.096$ & $1.234 \pm 0.029$ & ... \\

    \hline
     \hline
   \end{tabular}

% break table here    
    \begin{tabular}{l c c c l}
    \multicolumn{5}{c}{Table S3. (cont.)}\\
    \hline
    \hline
        Event Name & KMTNet Name & $\log{q}$ & $s$ & Reference \\
    \hline
    OB190249 & KB190109 & $-2.127 \pm 0.015$ & $0.543 \pm 0.010$ & \cite{Jung23AF8} \\
    & & $-2.127 \pm 0.016$ & $0.545 \pm 0.010$ & ... \\
    & & $-2.106 \pm 0.013$ & $1.777 \pm 0.014$ & ... \\
    & & $-2.107 \pm 0.013$ & $1.775 \pm 0.015$ & ... \\
    \hline
    OB190954$^{*}$ & KB193289 & $-2.044 \pm 0.042$ & $0.802 \pm 0.004$ & \cite{Han2021three} \\
    \hline
    OB181011$^{*}$ & KB182122 & $-2.007 \pm 0.011$ & $1.281 \pm 0.009$ & \cite{OB181011} \\
    & & $-2.034 \pm 0.010$ & $0.750 \pm 0.005$ & ... \\
    \hline
    OB181647 & KB182060 & $-2.001 \pm 0.028$ & $1.433 \pm 0.014$ & \cite{Gould22AF5} \\
    \hline
    OB190299 & KB192735 & $-1.998 \pm 0.028$ & $0.990 \pm 0.002$ & \cite{Han2021resonant} \\
        \hline
    OB190468$^{\dagger}$ & KB192696 & $-1.976 \pm 0.020$ & $0.717 \pm 0.007$ & \cite{OB190468} \\
    & & $-1.980 \pm 0.021$ & $1.379 \pm 0.012$ & ... \\
    \hline
    OB181011$^{\dagger}$ & KB182122 & $-1.824 \pm 0.018$ & $0.582 \pm 0.005$ & \cite{OB181011} \\
    & & $-1.817 \pm 0.017$ & $0.577 \pm 0.005$ & ... \\
    \hline
    \hline
\multicolumn{5}{l}{$\quad^*$Planet b in a two-planet system.}\\
\multicolumn{5}{l}{$\quad^\dagger$Planet c in a two-planet system.}\\
\hline
\hline
   \end{tabular}

\end{center}

%\FloatBarrier
\renewcommand\arraystretch{1.25}
\begin{center}
\begin{tabular}{p{0.075\textwidth} p{0.4\textwidth} p{0.525\textwidth}}
\multicolumn{3}{c}{\bf Table S4. Selection criteria for our sample of microlensing events}\\
\hline
\hline
 & Criteria & Description\\
\hline
Cut-1          &  $\Delta I > 0.1$ mag 
	& Require a minimum change in $I$-band magnitude between the baseline magnitude, $I_{\rm base}$, and peak magnitude, $I_{t_{0}}$, of the event: $\Delta I = I_{\rm base} - I_{t_{0}}$; rejects low-amplitude variables.\\
Cut-2          &  $u_{0} < 2 $ AND  &\\
	          &  $t_{\rm E} < 300~{\rm days}$  
	&   Reject extreme values for the $u_0$ and $t_{\rm E}$ parameters of the PSPL model, which usually correspond to light curves of non-microlensing events.\\
Cut-3          &  $\sigma(u_{0})/u_{0} < 0.3 $ AND \\
                  &  $\sigma(t_{\rm E})/t_{\rm E} < 0.3$        
	& Require the $u_0$ and $t_{\rm E}$ parameters of the PSPL model to be well-constrained; light curves that fail this cut usually correspond to non-microlensing events.\\
Cut-4          &  ($N_{\rm r} \geq 6$ AND $N_{\rm d} > 0$) OR   \\
		  & ($N_{\rm r} > 0$ AND $N_{\rm d} \geq 6$)                                
	& Require a minimum number of data points on both the rising and declining sides of the light curve, where $N_{\rm r}$ is the number of data points within the time interval $(t_{0} - t_{\rm E} < t < t_{0})$ and $N_{\rm d}$ is the number of data points within $(t_{0} < t < t_{0} + t_{\rm E}).$ \\
Cut-5          &  IF $0 < N_{\rm d} < 6$, \\
& $\quad N_{\rm r} \geq 6$ AND $N_{t_{\rm eff}} \geq 6$ \\
               &  IF $0 < N_{\rm r} < 6$, &\\
               & $\quad N_{\rm d} \geq 6$ AND $N_{t_{\rm eff}} \geq 6$
               & $N_{\rm t_{\rm eff}}$ is the number of data points in the time range $|t-t_{\rm 0}| < u_{0}{\times}t_{\rm E}$; rejects events that peak too close to or outside the beginning or end of the observing season.\\
Cut-6         &  $I_{\rm S} < 23.0$ mag                                                       
	& Require a maximum $I$-band source magnitude (equivalent to a minimum source brightness).\\
\hline
\hline
\end{tabular}
\end{center}

    \renewcommand\arraystretch{1.25}
    \begin{center}
    \begin{tabular}{c c c|c c c}
    \multicolumn{6}{p{\textwidth}}{{\bf Table S5. Microlensing planet mass-ratio function parameters.} The models were fitted to the data using a Markov Chain Monte Carlo. The best-fitting values are from the highest-likelihood sample. The full posterior probability distributions are summarized here as the median and 68\% confidence intervals.}\\
    \hline
    \hline
    \multicolumn{3}{c|}{Power-Law} & \multicolumn{3}{c}{Double-Gaussian} \\
    Parameter                    & Best-fitting value    & Posterior & 
    Parameter                    & Best-fitting value    & Posterior  \\
    \hline
    $\ln{\mathcal{L}}$         & $-28.22$ &  & 
    $\ln{\mathcal{L}}$         & $-16.90$ & \\
    $\mathcal{A}$                               & $0.175$  & $0.177^{+0.028}_{-0.026}$ & 
    $\mathcal{A}_1$                           & $0.475$  & $0.52^{+0.19}_{-0.13}$  \\
    $\gamma$                    & $-0.542$ & $-0.547^{+0.050}_{-0.049}$ & 
    $\mathcal{A}_2$                           & $0.057$  & $0.058^{+0.014}_{-0.012}$  \\
                                         & & & 
    $\log{q_{\rm peak,1}}$ & $-4.651$  & $-4.67^{+0.19}_{-0.33}$ \\
                                         & & & 
    $\log{q_{\rm peak,2}}$ & $-2.635$  & $-2.63^{+0.09}_{-0.11}$ \\
                                         & & &                              
    $n_1$                           & $-0.772$  & $-0.73^{+0.31}_{-0.41}$ \\
                                         & & & 
    $n_2$                           & $-1.727$  & $-1.78^{+0.54}_{-0.69}$ \\
    \hline
    \hline
    \end{tabular}\\
    \end{center}

\clearpage

\end{document}